\documentclass{aa}
\usepackage{epsfig}
\begin{document}
\title{Cool carbon stars in the halo: a new survey based on 2MASS
\thanks{Based on observations made at the European Southern Observatory, Chile
 (programs  67.B~0085AB, 69.B~0186A)
 and at the Haute Provence Observatory (France) operated
by the Centre National de Recherche Scientifique, together with data from the
2MASS project (University of Massachusetts and IPAC/Caltech, USA).}}
\author{N. Mauron\inst{1} \and  M. Azzopardi\inst{2} \and K. Gigoyan\inst{3}
\and T.R. Kendall\inst{4}} 
\offprints{N.Mauron}
\institute{ Groupe d'Astrophysique, UMR 5024 CNRS, Case CC72,
 Place Bataillon, F-34095 Montpellier Cedex 5, France\\
 \email{mauron@graal.univ-montp2.fr}
 \and 
  IAM, Observatoire de Marseille, 2 Place Le Verrier, 
F-13248 Marseille Cedex 4, France
\and 
378433 Byurakan Astrophysical Observatory \& Isaac 
 Newton Institute of Chile, Armenian Branch, Ashtarak d-ct, Armenia
\and
 Laboratoire d'Astrophysique, Observatoire de Grenoble, Universit\'e
Joseph Fourier, BP 53, 38041 Grenoble Cedex 9, France}

\date{Received xxx/  Accepted xxx}

\abstract{We present the first results
of a new survey for finding cool N-type carbon (C) stars in the
halo of the Galaxy.  Candidates were first selected  in  the 
2MASS Second Incremental Release  database 
with $JHK_s$ colours typical of red AGB C stars and $K_s < 13$, and
subsequently checked through medium resolution slit spectroscopy. 
We discovered 27 new C stars $plus$ one known previously and
 two similar objects in the Fornax and
Sculptor dwarf galaxies. We determine and discuss the properties of our sample, 
including optical and near-infrared colours, radial velocities, as well as
 $H\alpha$ emission and variability that are frequent, all  these characteristics
 being compatible with an AGB C-type
classification. Surprisingly, of the 30 studied objects, 8
 were found to have small but measurable
proper motions ($\mu$) in the USNO-B1.0 catalogue, 
ranging over $8 < \mu < 21$ mas\,yr$^{-1}$ and
opening the possibility that some objects could perhaps be dwarf carbon stars. Yet,
 a detailed analysis based on comparison with the sample of known carbon dwarfs 
leads us to consider these $\mu$ as incompatible with the broader picture suggested by
the other data taken as a whole. So, we adopt the view that all
objects are of AGB type, i.e. luminous and distant. 
Because the stream of Sagittarius dwarf galaxy is known to
be the dominant source of  luminous C stars in the halo, we chose to determine
distances for our sample by scaling them on the 26 known AGB C stars of the Sgr galaxy
 itself, which are found to be,  in the $K_s$-band, $\sim$ 0.5\,mag. less luminous than 
the average LMC C stars for a given $J-K_s$ colour. The obtained distances  of our halo stars 
range from 8 to 80\,kpc from the Sun. Then, examination of position and radial
velocities show that about half belong to the Sgr stream.
Our findings suggest that numerous AGB C stars remain to be discovered in the halo.
Long term K$_s$-band monitoring would be of great value to ascertain distance
estimates through the period-luminosity relation, because a large fraction
of our sample is probably made of Mira variables.
 \keywords{Stars:  carbon, surveys, galactic halo; Galaxy: stellar content} 
} 
\titlerunning{Halo carbon stars}
\authorrunning{ N.Mauron, M. Azzopardi, K. Gigoyan, T.R. Kendall}
\maketitle
\section{Introduction}

 Surveys of stellar populations located at high galactic latitude are
 important to characterize the halo and to understand how the Galaxy
 formed (see for example Majewski 1993, and references therein).
  Among the various types of stars that have been investigated 
 with this goal, the case of carbon (C) stars has been the subject of
much attention for some years. If such a C star is proven 
 to be in the asymptotic giant branch (AGB) evolutionary phase
  (as is often the case for cool C stars
 in the galactic disk), with an $R$-band magnitude of the order of 15, its
 high luminosity ($M_{R} \sim -3.5$) puts it as far as 
 50\,kpc from the Sun. Therefore, the luminous C stars constitute valuable
  probes of the distant halo (e.g. Bothun et al. 1991).
 Considerable efforts have been accomplished, and are still
 in progress, in order to find such faint high latitude carbon stars (FHLCs).
  These rare objects have been discovered using two main methods. The first is
   by exploiting Schmidt objective-prism plates where C stars have a conspicuous
 spectral appearance (MacAlpine \& Lewis 1978, Sanduleak \&  Pesch 1988, 
 Gigoyan et al. 2001, Christlieb et al. 2001). The second method uses a 
 preliminary selection of candidates
 with suitable photometric criteria, such as a very red $B$$-$$R$ colour index,
 as in the
 APM (Irwin 2000) survey of Totten \& Irwin (1998; hereafter TI98), or  
 multicolour properties as in the SLOAN carbon star survey of 
 Margon et al. (2002), with subsequent verification the of the carbon star nature 
 of these candidates by follow-up  spectroscopy. 
 
 One of the most striking results derived from these FHLC surveys, 
 especially from the APM one, was the fact that the tidal stream of the
  Sagittarius dwarf spheroidal galaxy
  (Sgr)  orbiting the Galaxy could be traced for the first time by 
 considering  the spatial and kinematical properties of the distant cool 
 C stars (Ibata et al. 2001a).
 The Sgr stream has now been detected through a number of other 
 methods, such as deep mapping in limited portions of sky of specific populations, 
 e.g.,  blue horizontal branch stars, metal poor K giants, turnoff stars,
 RR Lyr variables  (Dinescu et al. 2002, Dohm-Palmer et al. 2001,
  Kundu et al. 2002, Vivas et al. 2001, Martinez-Delgado et al. 2001,
  Newberg et al. 2002), or with 2MASS selected M giants over the whole
  sky (Majewski et al. 2003). Yet, there are several reasons to pursue the
  search for cool luminous C stars all over
 the high latitude sky. Firstly,  the  detection of this stream 
 with cool FHLCs currently involves only $\sim$ 40 stars, so that enlarging the sample
  is naturally desirable. Secondly, cool AGB C stars are 
 a population   of intermediate age, and consequently their spatial distribution 
 in the Sgr orbits  might provide some interesting information  on the history of the 
 merging process. A fraction of these AGB C stars may also be Mira 
 variables, and help to determine distances of these orbits through the
 period-luminosity relation. Finally, roughly
 half of the cool FHLC stars  do not belong to the Sgr stream; their origin
 has to be investigated and increasing the size of the sample studied may 
 possibly reveal other streams.
 
  However, in the search for FHLCs, one has to take into account that 
the C stars in general are of various types and with diverse
evolutionary origins (see Wallerstein \& Knapp 1998 for a review).
 Compared to the bright AGB stars,
 one family  consists of less luminous, warmer carbon-rich objects presently evolving as
 clump giants or located along the red giant branch, having accreted carbon 
 from a more evolved companion  (Knapp et al. 2001, Christlieb et al. 2001).
  Moreover, it is now well 
 established that a class of dwarf carbon stars exists (dCs; see e.g., 
 Dahn et al. 1977, Green 2000, Margon et al. 2002, Lowrance et al. 2003, and 
 references therein). 
 These dCs have very low luminosity, are located
 within a few hundred parsecs, have generally measurable proper 
 motions, and in fact are expected to outnumber the C stars of giant 
 type as observations
 probe successively fainter magnitudes (Margon  2003).

In this context, we report here on the first results of a new  systematic search for 
faint, red AGB C stars at high galactic latitude. Our survey is essentially
a near-infrared based survey, since  our candidates have been selected from the 2MASS 
Second Incremental Data Release point-source catalogue.
 About half of our $\sim$ 200 best candidates 
have now been observed spectroscopically, resulting in the discovery of 27 new
cool FHLCs
which are presented and analysed in this work.
After describing our selection method (Sect.~2), the spectroscopic observations 
are reported in Sect.~3. In Sect.~4, we examine  the various properties
of the sample, including  radial velocities, variability and proper
motions. These results are analysed in Sect.~5 together with a determination of
distances and examination of membership to the Sgr stream. 
The main conclusions are finally summarized in Sect.~6.

\section{Selection of candidates}

In order to find new FHLC stars, we first  considered all 
the FHLCs published in the literature and located at 
$|b| > 30\degr$.
After retrieving their $JHK_s$ photometry from the 2MASS  Second Incremental 
Data Release   point-source catalogue (available when this work started and
covering about half of the sky),  we plotted them in a colour-colour  $JHK_s$ 
diagram (Fig~1). Very similar diagrams, which inspired our search method,
 have been published  
by Totten, Irwin and Whitelock (2000; TIW), and Liebert et al. (2000).
It can be seen in Fig.~1 that the large majority of FHLCs have an $H-K_s$ colour 
of about 0.2.
These stars are relatively warm, presumably CH-type objects and come mainly 
from the Hamburg/ESO sample of Christlieb et al.  (2001). The cool N-type stars 
which we seek are located at $H-K_s$ larger than $\sim$ 0.3, and
 appear to form a relatively well defined locus up to $H-K_s \sim 1.1$,
although the number of objects is progressively decreasing. The width of this
locus is typically  $\sim$ 0.25 mag. This plot also suggests that 
the C star locus  extends  up to the two objects at 
$H-K_s \sim 1.6$, and such an extension is supported when one considers a similar 
diagram showing the LMC C stars listed the catalogue of Kontizas et al. (2001)
(see also e.g., Nikolaev \& Weinberg 2000, their Fig.~2).
At still redder colours, Fig.~1 also shows also two exceptionally cool N-type stars with  
$H-K_s \sim 2.0$. These are the very dusty C stars IRAS\,0846+1732, found by 
Cutri et al. (1989), for which $l=210^{\circ}$, $b=+35^{\circ}$, $J-H=2.37$, 
$H-K_s =2.01$ 
and $K_s=10.71$,  and IRAS\,03582+1819, found by Liebert et al. (2000), with
$l=210^{\circ}$, $b=-25^{\circ}$, $J-H=2.59$, $H-K_s=2.07$, and $K_s=9.26$.
The latter object is plotted in Fig.~1
despite having $|b| <  30^{\circ}$ because its height above the galactic plane is
estimated by Liebert et al. to be in the range 6--15\,kpc.

Our method for searching for cool C stars was therefore  
to select in 2MASS objects lying within a distance of $\sim$ 0.15\,mag 
to the median line formed by these template cases. In order to avoid 
a large number of ordinary (M-type) stars in our selection, we had also to set 
a limit on colours, e.g. $H-K_s > 0.4$, $J-H > 0.95$, meaning that
we naturally miss the numerous warm but less luminous giant C stars
that are much better selected by other techniques, e.g. through 
the SLOAN multicolour criteria. Concerning the limits in galactic latitude,
our nominal goal was to limit our search to $|b| > 30^{\circ}$.  However, 
we also considered with a lower priority candidates located down 
to  $|b| \sim 25^{\circ}$, especially if they showed an additional 
favourable property such as very red  $B-R$ or $J-K_s$ colours, and 6
 new C stars were found at these low latitudes (more details  in  Sect.~4). 
 

 Concerning the limit in brightness, our 
nominal limit was set by $K_s < 13$, which  corresponds to the 
rather large distance of $\sim$ 150\,kpc from the Sun, if one adopts as a basis  
the typical not too red LMC C stars  
with $J-K_s = 1.6$, which have a mean $K_s$ of $10.7$  ($\sigma = 0.4$).  

After selection in 2MASS which yielded $\sim$ 1200 objects, we  
excluded the objects that were  already known and catalogued 
in the SIMBAD database as M or C stars,
 young stellar objects, L dwarfs, galaxies or QSOs. We also excluded 
 objects with USNOC-A2.0 colour 
$B-R$ bluer than 1.5, when these $B$ and $R$ magnitudes are provided by 
the 2MASS database. This is justified by the fact that many M stars and galaxies
are excluded by this criterion, while  N-type stars are 
expected to be much redder than this limit and have generally $B-R \sim 3$. 
Eventually, we found 6 new C stars with $2 <B-R < 3$ and 2 with $B-R$ = 1.9 
and 1.6  (see below).  
Inspection of POSS plates was also systematically used for further sample cleaning, 
and numerous supplementary cases of faint, contaminating galaxies were discarded. 
In addition, the objective-prism plates of the First Byurakan Survey were
examined by one of us (K.G.) for relatively bright candidates located in the
zones covered by this survey, and this allowed the elimination of a number of further
M-type stars. This process resulted in a list of $\sim$ 200 best   
candidates for which slit spectroscopy follow-up was begun.

\section{Observations}

\begin{table}
        \caption[]{Journal of Observations }
\begin{center}
        \begin{tabular}{lll}
        \noalign{\smallskip}
        \hline
	\hline
        \noalign{\smallskip}
 Run \# and dates & Site  &  \# of observed Objects\\
\noalign{\smallskip}
\hline
\noalign{\smallskip}
1~~  2001 Mar 26 - Mar 30 & OHP  & 2, 3, 5 to 10, 14\\
2~~  2001 Mar 31 - Apr 01 & ESO  &4, 7, 11, 12, 13\\
3~~  2001 Sep 09 - Sep 13 & ESO  &1, 16, 17, 19, 22 to 28\\
4~~  2001 Oct 17 - Oct 22 & OHP  &15\\
5~~  2002 Fev 14 - Fev 18 & OHP  &no observations (clouds)\\
6~~  2002 Aug 29 - Sep 03 & ESO  &17, 18, 20, 21, 29, 30\\
\noalign{\smallskip}
\hline
\end{tabular}
\end{center}
\end{table}


 The observations were carried out  with the
 193\,cm telescope at  Haute-Provence Observatory in France (OHP) and with
 the Danish 1.54\,m telescope at the European Southern Observatory (ESO) 
 in Chile. A journal of observations is given in Table 1, indicating the
 dates of the observing runs and the objects observed in each run.
 Because of clouds, no observations were done during Run 5, 
 mentioned here for completeness.
 
  At OHP we used the CARELEC spectrograph and its  1200 lines\,mm$^{-1}$ grating 
  blazed in the red to obtain  a dispersion of 0.45\,\AA\,pixel$^{-1}$ and 
 to cover the 5700\,\AA -- 6600\,\AA~ region. 
 The detector is an EEV 42-20 CCD chip with 2048$\times$1024 pixels of 13.5\,$\mu$m.
 Due to poor seeing, the slit width had to be set to 2$''$, and the resulting velocity
 resolution of the spectra is  $\sim$ 90\,km\,s$^{-1}$. 
 
 At ESO, we used the DFOSC focal reducer which permits both
 direct imaging and slit spectroscopy. After a 2\,min image generally taken
 in the R Bessel filter for source identification, the spectrum was
 obtained through grism \#8 which provides a range of 5800 to 8400\,\AA~ and 
 a dispersion of 1.2\,\AA\,pixel$^{-1}$ on the detector, a 
 2148$\times$4096 EEV/MAT CCD with 15\,$\mu$m pixels (half of the CCD area 
 is not used due to the reducer design). 
 The slit width was 1.5$''$ and the velocity resolution $\sim$ 120\,km\,s$^{-1}$.

\begin{figure*}
\resizebox{\hsize}{!}{
{\rotatebox{-90}{\includegraphics{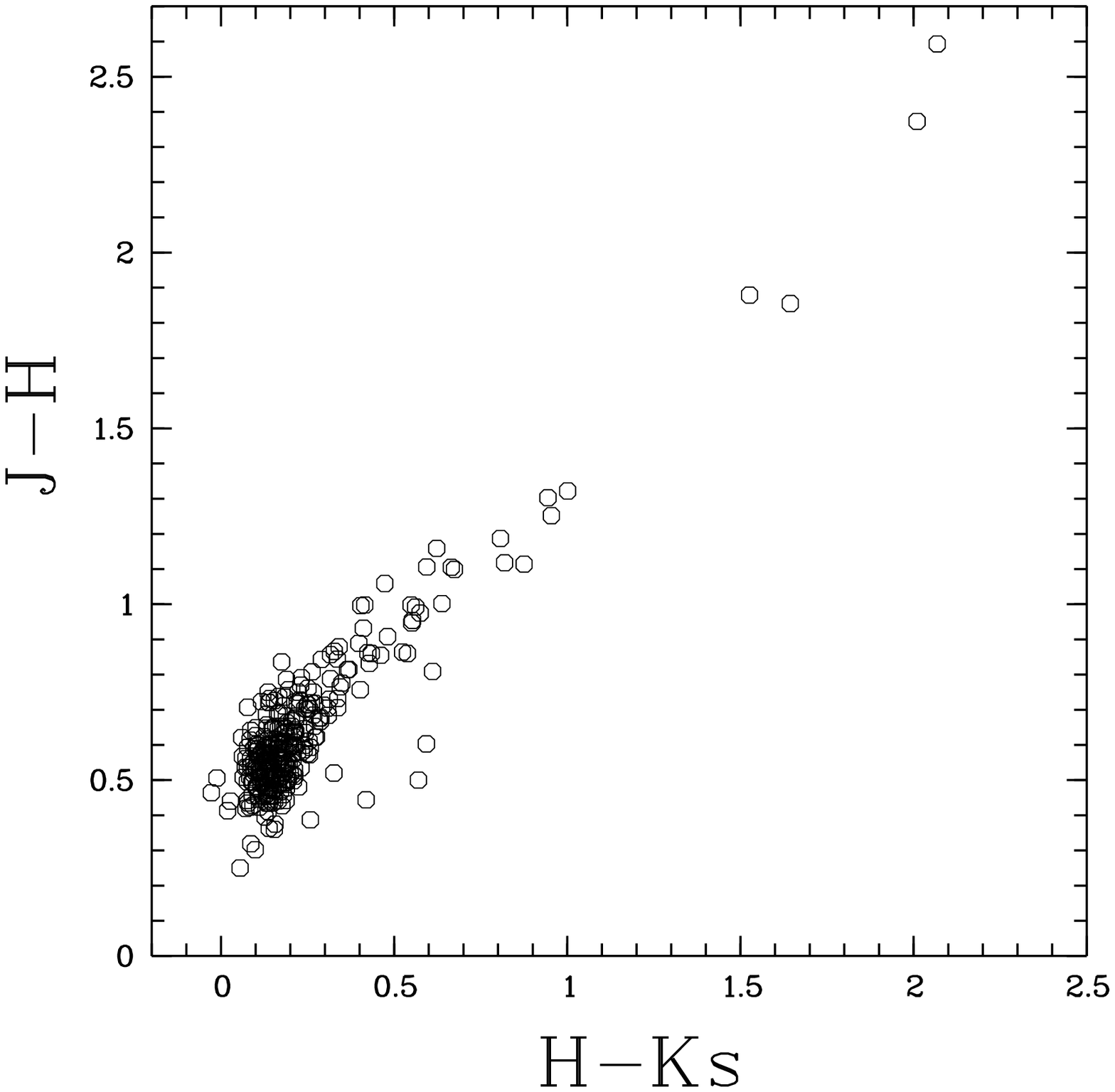}}}
{\rotatebox{-90}{\includegraphics{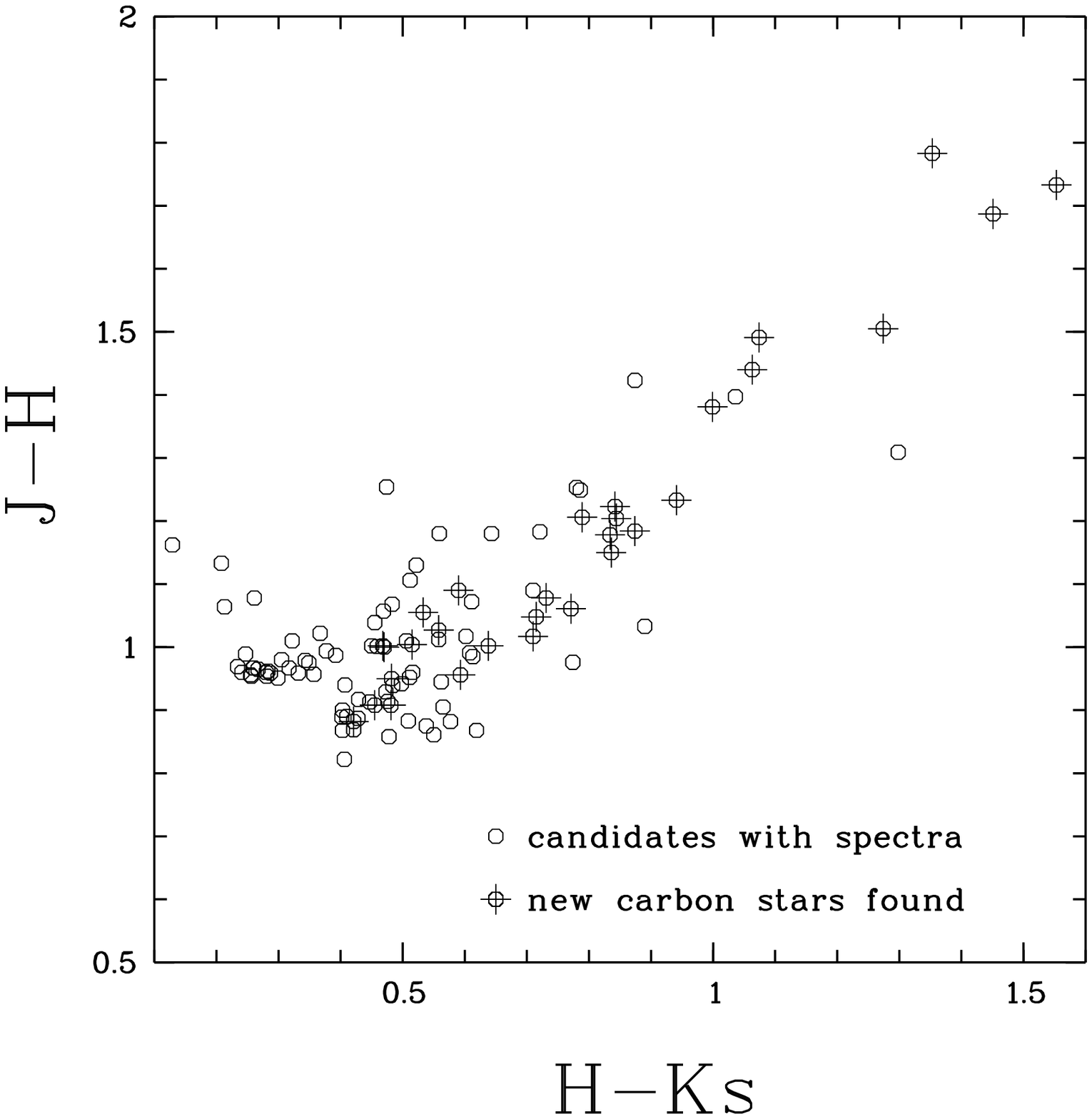}}}
}

\caption[]{{\it Left panel}: Colour-colour diagram of known carbon stars with 2MASS photometry and
located at  high galactic latitude ($|b| > 30^{\circ}$);
{\it Right panel}: Colour-colour diagram of the targets for which
slit spectroscopy have been obtained (circles). The new carbon stars
found in this work are indicated by an overplotted $+$ sign. Note that the abscissa and
ordinate scales differ in the two panels.}
\label{fig1.ps}
\end{figure*}

\begin{table*}
	\caption[]{List of discovered faint cool  halo carbon stars. Coordinates
	$\alpha$ (in h., min., sec.) and $\delta$ (in deg., min, sec.) are from 2MASS.
	$l$, $b$ are in degrees. $B$ \& $R$ in mag. are from USNO-A2.0 
	($\pm$ 0.4 mag. approximatively), 
	except for objects with a Note. $J H K_s$ are from the 2MASS 2nd Incr. Release
	database (in mag.; errors $\pm$ 0.02-0.03 mag. or better)}
	\begin{flushleft}
	\begin{tabular}{lccrrrrrrrrrr}
	\noalign{\smallskip}
	\hline
	\hline
	\noalign{\smallskip}
No.&  $\alpha$(J2000) & $\delta$(J2000) & $l$~~~~ & $b$~~ & $B$ & $R$~~ & $B$-$R$ & $J$~~ & $H$~~ & $K_s$~~ & $J$-$K_s$&Notes\\   
	\noalign{\smallskip}
	\hline
        \noalign{\smallskip}

  01& 02 11 30.866& $-$03 49  43.85& 165.80& $-$59.86&  18.5 & 14.9 & 3.6 & 12.034 & 11.007 & 10.449 & 1.585&\\
  02& 09 13 31.865& $+$19 34  22.64& 209.31& $+$39.79&  16.9 & 11.5 & 5.4 &  9.043 &  7.953 &  7.363 & 1.680&\\
  03& 09 15 05.206& $+$19 17  37.89& 209.81& $+$40.04&  16.8 & 12.6 & 4.2 & 10.203 &  9.186 &  8.476 & 1.727&\\
  04& 10 15 25.934& $-$02 04  31.84& 244.49& $+$42.43&  20.9 & 16.3 & 4.6 & 14.045 & 12.861 & 11.987 & 2.058&(1)\\
  05& 10 59 23.839& $+$39 44  05.60& 177.25& $+$63.60&  16.9 & 13.3 & 3.6 & 11.082 & 10.080 &  9.442 & 1.640&\\
  06& 11 09 59.686& $-$21 22  01.15& 273.53& $+$35.65&  14.2 & 10.5 & 3.7 &  8.390 &  7.482 &  7.001 & 1.389&\\
  07& 11 17 19.005& $-$17 29  15.39& 273.18& $+$39.88&  19.4 & 16.5 & 2.9 & 13.464 & 12.403 & 11.632 & 1.832&\\
  08& 12 09 25.022& $+$15 16  18.49& 261.33& $+$74.64&  17.8 & 14.2 & 3.6 & 11.185 & 10.277 &  9.822 & 1.363&\\
  09& 12 49 04.767& $+$13 20  35.51& 300.53& $+$76.20&  19.3 & 14.5 & 4.8 & 12.606 & 11.604 & 11.136 & 1.470&\\
  10& 13 56 02.371& $-$01 36  26.20& 333.93& $+$57.32&  19.4 & 15.5 & 3.9 & 12.915 & 11.860 & 11.327 & 1.588&\\
  11& 13 59 20.636& $-$30 23  39.48& 319.88& $+$30.23&    -  & 19.8 &  -  & 14.577 & 13.072 & 11.798 & 2.779&(2)\\
  12& 15 01 06.923& $-$05 31  38.70& 351.63& $+$44.74&  20.4 & 16.8 & 3.6 & 13.571 & 12.348 & 11.506 & 2.065&(3)\\
  13& 15 15 11.063& $-$13 32  27.93& 348.10& $+$36.43&  18.7 & 17.1 & 1.6 & 12.594 & 11.516 & 10.785 & 1.809&(4)\\
  14& 15 58 42.227& $+$18 52  46.86&  32.33& $+$46.37&  17.5 & 14.6 & 2.9 & 12.269 & 11.387 & 10.966 & 1.303&\\
  15& 17 28 25.766& $+$70 08  29.93& 100.83& $+$32.41&  18.1 & 13.9 & 4.2 & 11.551 & 10.111 &  9.048 & 2.503&\\ 
  16& 19 42 19.018& $-$35 19  37.69&   4.40& $-$25.06&  18.6 & 16.7 & 1.9 & 12.633 & 11.142 & 10.068 & 2.565&\\
  17& 19 42 21.315& $-$32 11  04.19&   7.70& $-$24.13&  19.0 & 14.7 & 4.3 & 11.967 & 10.817 &  9.981 & 1.986&\\
  18& 19 48 50.653& $-$30 58  31.92&   9.43& $-$25.08&   -   & 17.7 &  -  & 12.998 & 11.215 &  9.862 & 3.136&(5)\\
  19& 19 53 30.172& $-$38 35  59.40&   1.52& $-$28.07&  19.2 & 13.9 & 5.3 & 11.292 & 10.088 &  9.244 & 2.048&\\
  20& 20 13 19.435& $-$23 41  44.26&  19.07& $-$27.92&  14.4 & 11.7 & 2.7 &  9.541 &  8.591 &  8.109 & 1.432&\\
  21& 20 20 27.661& $-$14 49  27.10&  29.05& $-$26.26&  19.4 & 14.7 & 3.7 & 11.849 & 10.162 &  8.711 & 3.138&(6)\\
  22& 20 54 54.551& $-$28 28  56.73&  16.76& $-$38.23&  18.6 & 14.3 & 4.3 & 12.407 & 11.451 & 10.858 & 1.549&\\
  23& 22 05 14.590& $+$00 08  46.06&  60.31& $-$41.67&  18.1 & 14.3 & 3.8 & 10.709 &  9.503 &  8.714 & 1.995&\\
  24& 22 06 53.669& $-$25 06  28.28&  26.55& $-$53.17&  18.0 & 15.1 & 2.9 & 10.934 &  9.756 &  8.922 & 2.012&\\
  25& 22 17 09.923& $-$26 07  03.35&  25.64& $-$55.64&  18.4 & 15.4 & 3.0 & 11.056 &  9.823 &  8.882 & 2.174&\\
  26& 23 17 21.087& $-$24 11  42.41&  35.54& $-$68.63&  17.8 & 15.5 & 2.3 & 13.750 & 12.750 & 12.280 & 1.470&\\
  27& 23 19 35.533& $-$18 56  23.79&  49.28& $-$67.38&  17.5 & 14.8 & 2.7 & 11.477 & 10.473 &  9.958 & 1.519&\\
  28& 23 25 31.394& $-$30 10  56.06&  18.64& $-$70.94&  18.4 & 14.6 & 3.8 & 13.409 & 12.028 & 11.029 & 2.380&\\
\multicolumn{13}{c}{Two carbon stars in the direction of Sculptor (\#29) and in Fornax (\#30)}\\
  29& 00 59 53.680& $-$33 38  30.77& 287.82& $-$83.24&   -   & 20.2 &  -  & 14.877 & 13.144 & 11.591 & 3.286&(7)\\
  30& 02 41 03.550& $-$34 48  05.34& 237.84& $-$65.37&  23.3 & 18.3 & 5.0 & 14.445 & 13.397 & 12.682 & 1.763&(8)\\
   \noalign{\smallskip}
   \hline
\end{tabular}
\end{flushleft}

{\small Notes:

(1)  $B$ \& $R$ are $B_2$ \& $R_2$ from USNOC-B1.0 in which $R_1$=18.5; in the APM 
catalogue,  one finds $R$=18.3 and no data for $B$

(2) $R$ is $R_2$ from USNOC-B1.0 in which no other data in $R$ or $B$ are given;
 in APM, $R$=20.25 and no data in $B$

(3) $B$ \& $R$ are $B_2$ \& $R_2$ from USNOC-B1.0 in which $R_1$=16.0;
in  APM, the object is blended with neighbours

(4) $B$ \& $R$ are $B_2$ \& $R_2$ from USNOC-B1.0 in which $R_1$=16.7; in APM,
$R$=17.1 $B$=18.35

(5) $R$ is $R_2$ from USNOC-B1.0 in which $R_1$=15.8 but no data in $B$ is given; in APM,
$R=18.2$ and no data in $B$

(6) $B$ \& $R$ are $B_2$ \& $R_2$ from USNOC-B1.0 in which $R_1$=16.2; no data
in the APM catalogue for this position ($|b|$ is too low)

(7) $R$ is $R_2$ from USNOC-B1.0 in which no other data in $R$ or $B$ are given; in APM,
 $R$=20.3 and no data in $B$; membership to Sculptor requires supplementary
 observations and radial velocity determination.
 
(8) $B$ \& $R$ are $B_2$ \& $R_2$ from USNOC-B1.0 in which $R_1$=16.9; in APM,
$R=18.3$ and no data in $B$; this star was previously identified as probable C star
by Demers et al (2002) on the basis of its near-infrared photometry
 (\#25 in their Table 1)}

\end{table*}

\section{Results}

 In our list of $\sim 200$  best candidates, slit 
 spectroscopy has so far been secured for 97 of them: 30 were found to be C stars, 
 including one that is member of the Fornax dwarf galaxy (\#30) and one located
 in the direction of the Sculptor dwarf galaxy (\#29) (see Table 2). The 67 other objects 
 (not  C stars) were found to be mainly M-type giants  and will be
 the subject of future work.

  The last two objects (\#29 and \#30) were
 under consideration as interesting comparison objects.   No radial velocity could be
 determined by us for \#29, and its membership to Sculptor needs further observations to be 
 proven.  This object is considerably redder ($J-K_s$ = 3.3) than the other C stars
 known in Sculptor (Azzopardi et al. 1986, Aaronson \& Olszewski 1987) for which
 $J-K_s$ is between 0.8 and 1.12. It appears very faint in the R-band POSS-II image, is
  invisible on blue plates, but is well seen in the I-band UKST digitized image.
  Concerning Object \#30, this star  was already noted by Demers et al. (2002) as a probable
 Fornax carbon star based on its 2MASS near-infrared magnitudes and colors: our spectrum
 confirms its carbon nature and proves its membership through radial velocity determination.

 All of the C stars found have $ K_s < 12.3$,
 with the exception of the Fornax C star at $K_s = 12.68$. We also observed a small 
 supplementary list of 9 faint ($13 < K_s < 14$) objects, none of which 
 were found to be C stars. The $JHK_s$ colour-colour diagram of the observed targets 
 is shown in Fig.~1 (right panel). During our survey, we also found ten L-type dwarfs,
 all with $12 < K_s < 14$, including seven which were not previously known (the three known
 cases had escaped our attention in the selection process).
 The discovery of these new L-dwarfs is reported in Kendall et al. (2003),
 and in the following, we focus on the new C stars and their properties.

\subsection{General properties of the sample}

Table 2 lists the 30 C stars found, their coordinates and some photometric data. 
Finding charts are not presented here, because all stars are near-infrared (NIR)
 bright and very red,
 and can be identified unambiguously in the 2MASS survey images, 
 and also in the POSS, ESO or UKST digitized images 
(note that the very red Object\#29 is clearly visible only in the IV-N SERC-I
 digitized  plate).
 
Here we shall ignore the last two C stars (\#29 and \#30) which are in the Fornax and 
Sculptor galaxies respectively. It can be seen that of the 28 remaining 
objects, 22 have been found 
at $|b| > 30^{\circ}$. One star, \#5, was erroneously rediscovered and was known as
FBS 1056+399 or APM 1056+4000 (Gigoyan et al. 2001, TI98). Its 7500--8000\,\AA\, spectrum
is in Gigoyan et al. (2001), and a 5700--6600\,\AA\, spectrum has been 
obtained here: it shows H$\alpha$ in emission and this new 
spectrum was used to derive an independent radial velocity
measurement which is in very good agreement with that of Gigoyan et al. (2001).


In Table~2, the columns $B$ and $R$, and corresponding $B-R$ index 
are from the USNO A2.0 catalogue, except 
for 8 objects which are not present in this catalogue, in
 which case the data from USNO-B1.0 are
given (see the Notes of Table 2 for details). These $B$ and $R$ magnitudes
 provide only approximate optical photometry with probable errors  of $\sim$\,0.4\,\,mag, 
and should also be considered with caution since many objects are clearly variable,  
and  objects at $|b| < 30^{\circ}$  may also suffer some interstellar absorption (see below).
However it is interesting to note that the magnitude range in $R$ is between 
10.5 and  19.8, and the median in $R$ is 14.7. For comparison, the carbon stars of
the APM survey (see Table 3 of TI98, with 41 stars labelled ``APM'') 
have an $R$ range 10.0--18.0 and a median of 13.6\,mag. The median of our sample 
14.7 corresponds to a distance of $\sim$ 44\,kpc if one adopts the absolute magnitude 
of $M_R$ = $-$3.5 considered by TI98, and if no  circumstellar or interstellar 
absorption in the red ($A_R$) is assumed (if $A_R$=1 mag., one finds 27\,kpc).

The columns $J, H, K_s, J-K_s $ are from the 2MASS Second Incremental release. 
 The typical errors on this photometry are $\sim$ 0.03\,mag or better.
 All C stars of our sample 
have $J-K_s > 1.3$, which is largely due to our selection criteria excluding
objects with $J-H < 0.95$ and $H-K_s < 0.4$. Therefore, they are distinctly
redder in $J-K_s$  than most of the numerous warm C stars of the Hamburg-ESO survey.
 The colour-colour diagram
of Fig.~1 also shows that no candidate redder than $J-K_s \sim 3.4$ was observed,
essentially because C stars  or candidates 
redder than this are very rare at high $b$.

\subsection {Spectra}

For clarity of the text, the atlas of all the spectra is shown in Appendix A.
A detailed study of these spectra will be performed in a future paper, 
and only a few remarks will be made here.

First, all ESO spectra display a strong rising flux between 6000 and 7800\,\AA\,
 with a flux ratio of about 2 to 15. The OHP spectra have a too small
domain to be considered similarly. This slope is clearly larger for our
objects than for the warmer giant or dwarf SLOAN C stars shown by Margon et al.
(2002), for which
the 6000--7800\,\AA\, flux distribution is nearly flat. The cool APM stars and the two
N-type C stars SDSS J144631.1-005500 and J1227400-002751 
(previously known as APM\,1225-0011) shown in
Margon et al. have spectra very similar with ours.

One notes also that H$\alpha$ is in emission in 13 of our 28 halo objects, 
i.e., 46\%. This fraction is higher but comparable to the result of Maizels \& Morris
 (1990) who surveyed 37 galactic ``bright C stars'' (presumably of AGB type, but no
details are given  on the observed stars or their selection method) 
and found H$\alpha$ emission
in 14 of them (38\%). In the APM survey, examination of the spectra in Fig. 5 of
TI98 indicates that 6 of 20 N-type stars have H$\alpha$ emission (30\%), 
and 1 of 8 CH-type stars (12\%). In contrast, among Margon et al.'s warm C stars,
 5 out of 39 have H$\alpha$ emission (13\%). Therefore the high fraction of 
 H$\alpha$ emission  in our sample and the above comparisons suggest that 
 most of our stars are
pulsating AGB stars (with a shock wave being the cause of the H$\alpha$ emission).

 
  \subsection{Radial velocities}
 
During each run, we observed several times a small number of template 
carbon stars  with known radial velocities (see Table 3). In the 
following, we shall call these stars ``radial velocity standard'', 
although, for several reasons,  they cannot be considered as classical 
standard stars with stable and accurately established 
velocities. Firstly, it is well known that the photospheric radial velocity
of cool carbon stars may vary with amplitudes of the order 
of $\sim$ 10\,km\,s$^{-1}$. Secondly, the number of independent radial velocities 
available in the  literature for a given star is often small. 
Thirdly, the literature values are  occasionally very discrepant: 
for example, in Table 4 of TI98, it is found that  APM\,0102-0556 \& 
APM\,0911+3341 have published values differing by as much as 50\,km\,s$^{-1}$.

Therefore, the published radial velocities of these standards have been
considered  as a first approximation. For each run considered separately,
we have cross-correlated  the spectra of each standard with all the other
standards observed during the run, and determined best fit radial
velocities by minimizing the differences between our data and the published 
values. The results,  listed in Table 3, appear fairly  consistent, especially 
when  comparing the velocities of the standards common to several runs and
taking into account the velocity resolution of our experiments.
The global rms scatter ($1\sigma$) of the residuals between fitted and 
published values is 12\,km\,s$^{-1}$.

\begin{table*}
        \caption[]{Heliocentric radial velocities for template carbon stars.  
 In columns labelled Run~1, Run~2, ..., Run~6 are listed	
the best fitted radial velocities determined for each run, $v_{\rm fit}$, in 
km\,s$^{-1}$. The column $v_{\rm pub}$ is the value taken from the literature and
 adopted as a first approximation for the fits (see text and Notes).
  The column  $ < v_{\rm fit} - v_{\rm pub} >$ gives the average difference over the runs
  (in km\,s$^{-1}$). }

	\begin{center}
     
        \begin{tabular}{lrrrrrrcc}
        \noalign{\smallskip}
        \hline
	\hline
        \noalign{\smallskip}
 Star & Run 1   & Run 2 &  Run 3& Run 4&  Run 6 & $v_{\rm pub}$ & $ < v_{\rm fit} - v_{\rm pub} >$ & 
 Note  on $v_{\rm pub}$\\
\noalign{\smallskip}
\hline
\noalign{\smallskip}
TW\,Oph           & +21    &  +28  &  +10  &     &       &  +14  & \,\,$+$6      & (1)\\
APM\,1406+0520    & $-$25  &  $-$37&       &     &       &  $-$21& $-$10   & (2)\\ 
HR\,3541 = X\,Cnc  & $-$4  &       &       &     &       &  $-$3 & \,\,$-$1   & (3)\\
APM\,0915$-$0327  &        &  +95  &       &     &       &   +79 & $+$16     & (2)\\
APM\,0123+1233    &        &       &$-$324 &     &       & $-$302& $-$22   & (2)\\
APM\,0418+0122    &        &       &  +19  &  +27&       &   +33 & $-$10   & (2)\\
APM\,2225$-$1401  &        &       &$-103$ &     & $-118$& $-113$& \,\,$+$3      & (2)\\      
APM\,0222$-$1337  &        &       &$ -7$  &     & $-24$ & $-27$ & $+$11    & (2)\\
RV\,Aqr           &        &       &$ +6$  & $-4$&       & $-10 $& $+$11    & (2)\\ 
HD\,16115         &        &       &$ -6$  &     &  +11  & $+4$  & \,\,$-$1  & (4)\\
APM\,2213$-$0017  &        &       &       &     & $-49$ & $-44$ & \,\,$-$5      & (2)\\
APM\,2111+0010    &        &       &       &     & $-195$& $-208$& $+$13     & (2)\\
\noalign{\smallskip}
\hline
\end{tabular}
\end{center}

Notes: 

(1) $v_{\rm pub}$ is from the SIMBAD database; 
from  the CO millimeter observations listed by Loup et al. (1993), a  center of mass
heliocentric velocity of +15\,km\,s$^{-1}$  is derived and is in very good agreement. 

(2) $v_{\rm pub}$ are from TI98;  quoted
uncertainties are $\leq$ 7\,km\,s$^{-1}$; 

(3) TI98 indicate $v_{\rm pub}$= $-$1\,km\,s$^{-1}$ with 
$\sigma$ = 12\,km\,s$^{-1}$ (2 measurements); 
from the data in Loup et al. catalogue, one derives $-$6\,km\,s$^{-1}$ (heliocentric), and
$v_{\rm pub}$ = $-$3\,km\,s$^{-1}$ was adopted;

(4) we adopted $v_{\rm pub}$ = +4\,km\,s$^{-1}$ from TI98, who give 
$\sigma$ = 1\,km\,s$^{-1}$ (3 measurements), but  
$v = +16 \pm 5$\,km\,s$^{-1}$ is given in SIMBAD. 

\end{table*}

Then, the spectrum of each program carbon star was correlated with
the spectra of the standards observed for the same run, and the velocities 
so obtained were averaged (details on the cross-correlation technique
can be found in, e.g., TI98). In addition to the internal consistency
provided by the standards, a further check is provided by two objects 
that were observed during two runs: for Object \#7, we found
$v_{\rm helio} = +339$ and  $+$345\,km\,s$^{-1}$ from Run 1 and 2 respectively; 
for Object \#17, we found $v_{\rm helio} = +135$ and $+$127\,km\,s$^{-1}$ from Run 2 and 6,
respectively. A final independent check on our velocity scale 
is provided by Object \#29 in Fornax, for which
we find $v_{\rm helio} = +$40\,km\,s$^{-1}$, in fair agreement with
the mean velocity of this galaxy $v_{\rm helio} = +$53\,km\,s$^{-1}$
(van den Bergh 2002), since the difference of 13\,km\,s$^{-1}$ represents
1.1$\sigma$. In conclusion, we estimate that the uncertainty
on our radial velocities is $\sim$~12\,km\,s$^{-1}$ (1$\sigma$).

\subsection {Proper motions}

Thanks for the recent release of the USNO-B1.0 catalogue (Monet et al. 2003),
information on proper motions  is available for all the objects.
For 22 objects, the proper motion is found to be null, while
for 8 objects, a measurable proper motion  is provided. For these 8 objects, 
details are listed in Table 4, including the time interval between first and last plates
on which the objects were detected, and the total motion probability as provided
by  USNO-B1.0.

We checked whether the objects for which a zero proper motion is given
have been correctly observed in the scanned surveys. Because most of 
the objects are at a declination larger than $\sim -30^{\circ}$ and
bright enough ($R$ in the range 14 to 18), 
they have been well imaged in the first POSS-I survey: they are generally
 well  detected in  red plates and very often also in blue plates of POSS-I, and they
are also well detected in subsequent red or near-infrared surveys 
(POSS-II, SERC, AAO, ESO), ensuring  time baselines which are of the order of
30 to 50 years.

\begin{table}
        \caption[]{Data from the USNO-B1.0 database 
	for objects with not null proper motions, where $\Delta t$ is
	the epoch interval, N is the number of detections on the scanned plates, and
	Probab. is the total motion probability}
\begin{center}
        \begin{tabular}{crrrrrr}
        \noalign{\smallskip}
        \hline
	\hline
        \noalign{\smallskip}
 Object  & $\mu_{\alpha}$ cos $\delta$ & $\mu_{\delta}$  &  $\Delta t$ & N & Probab. \\
  \#     &(mas/yr)                     & (mas/yr)        & (yrs)       & & \\
\noalign{\smallskip}
\hline
\noalign{\smallskip}
10 & $+24\pm3 $    & $2\pm3$  & 45 & 5 &0.9 & \\
17 & $-12\pm10$    & $6\pm1$  & 35 & 5 &0.7 & \\
18 & $-6\pm3$      & $-4\pm0$ & 33 & 3 &0.7 & \\
20 & $ -6\pm3 $    & $14\pm5$ & 38 & 5 &0.8 & \\
22 & $-16\pm2 $    & $0\pm3 $ & 36 & 4 &0.9 & \\
24 & $-8\pm2  $    & $6\pm4$  & 37 & 5 &0.9 & \\
25 & $-12\pm2$     & $2\pm5$  & 41 & 4 &0.9 &\\
26 & $-2\pm3$      & $8\pm3$  & 41 & 5 &0.9 &\\

\noalign{\smallskip}
\hline
\end{tabular}
\end{center}
\end{table}

\subsection{Variability}
All objects were examined for variability in available digitized sky 
surveys with 5$'$ size images retrieved from the 
USNO web site\footnote{www.nofs.navy.mil/data/fchpix}.
Practically, we compared by eye the appearance of our C stars with 
neighbouring stars of similar brightness on these survey images and on our ESO
CCD Bessel-R band images, when possible. One difficulty in doing so is
 that the photographic survey plates have been exposed
with a variety of  emulsions and filters. This complicates the comparison, 
especially because  the C stars are very red and have relatively steep 
spectra compared to neighbouring field stars. For example, the 
available red plates for a given field may have the following various emulsions 
and filters: 103aE + RP2444 (POSS I), IIIaF + RG 610-3 (POSS II), IIIaF + OG590 (AAO-R) 
or  IIIaF + RG 630 (ESO-R). Consequently, when examining these plates, we 
kept in mind possible bandpass effects  and concluded the existence of variability 
only when the evidence was very strong. For several objects, there are pairs 
of plates with identical emulsion/filter combinations with exposures 
taken at quite different dates, in which case variability is much better
assessed. In such cases, we noted 'var 2r' or 'var 2b', corresponding
to a pair of red or blue plates, respectively. In one case (Object \#17),
two CCD Bessel $R$ images were obtained during 2 different runs,
 and differential photometric analysis
of the frames very clearly establishes variability, by 0.28 $\pm$ 0.02\,mag over
 a period of $\sim$ 1\,year. 
 
As can be seen in Table 5, of the 28 FHLC stars in our sample,
variability is found for 11 objects. Since our method is sensible to only
large variations, it is most probable that a larger fraction of
objects is actually variable. The Sculptor C star (\#29) is also found to be
variable. Concerning the Fornax C star(\#30), the evidence for variability was 
not conclusive from the examined plates, but variability has 
been proven by the CCD optical imaging survey 
of Bersier \& Wood (2002).

\begin{table*}
        \caption[]{Properties of the halo carbon stars. Quantities $l$, $b$, $R$, $K_s$, $J-K_s$ are
	repeated from Table 2 for information.  Variability derived from examination of various 
	plate surveys is indicated by ``var'' (see text). ``H$\alpha$'' means H$\alpha$ in emission. $\mu$ is
	the USNO-B1.0 proper motion in mas\,yr$^{-1}$. $A_R$ is the adopted
	 R-band galactic extinction as given by Schlegel et al. maps, in mag. $M_{Ks}$ is the
	 adopted $K_s$-band absolute magnitude (see text).  The distances $d_R$ and $d_K$ are
	 in kpc. $g^*_{\rm A-type}$ is the Sloan $g$ mag. of an imaginary ``A-type'' population with
	 $M_{g}$=1.0 if it was at the distance $d_K$ from the Sun; $g^*_{\rm A-type}$ is used
	 to evaluate the membership of the  sample C star to the Sgr steam. This membership
	 is  given in  the last column by $yes$ or $no$ (see text). For objects \#29 and \#30, 
	 $M_{Ks}$ and distances based on Sgr C stars templates are provided here only for information
	 (see text)}
        \begin{flushleft}
        \begin{tabular}{lccrrrlrrrrrrrrrr}
        \noalign{\smallskip}
        \hline
        \hline
        \noalign{\smallskip}
No.&  $l$ & $b$ & $R$ & $K_s$ & $J-K_s$ & var & H$\alpha$& $\mu$ & $v_{\rm helio}$&  $A_R$ & $M_{Ks}$ &
 $d_R$ & $d_K$& $g^*_{\rm A-type}$ & Sgr~?\\   
        \noalign{\smallskip}
        \hline
        \noalign{\smallskip}

  01& 165.80& $-$59.86&  14.9 &  10.449 & 1.585&        &     & 0 & $-$104 & 0.06 &$-7.35$& 39 & 36&18.8&yes :&\\
  02& 209.31& $+$39.79&  11.5 &   7.363 & 1.680&        & H$\alpha$  & 0 &  $-$46 & 0.12 &$-7.45$&  8 &  9&15.8&  no&\\
  03& 209.81& $+$40.04&  12.6 &   8.476 & 1.727&        & H$\alpha$  & 0 &  $-$14 & 0.11 &$-7.50$& 13 & 16&17.0&  no&\\ 
  04& 244.49& $+$42.43&  16.3 &  11.987 & 2.058& var    &     & 0 & +202   & 0.09 &$-7.55$& 73 & 81&20.4&  no&\\
  05& 177.25& $+$63.60&  13.3 &   9.442 & 1.640&        & H$\alpha$  & 0 & $-$162 & 0.04 &$-7.40$& 19 & 23&17.8& no :&\\ 
  06& 273.53& $+$35.65&  10.5 &   7.001 & 1.389&        & H$\alpha$  & 0 & +124   & 0.11 &$-7.05$&  5 &  6&14.9&  no&\\
  07& 273.18& $+$39.88&  16.5 &  11.632 & 1.832& var 2b & H$\alpha$  & 0 & +342   & 0.12 &$-7.60$& 79 & 71&20.3&  no&\\
  08& 261.33& $+$74.64&  14.2 &   9.822 & 1.363&        & H$\alpha$  & 0 &  $-$27 & 0.09 &$-7.00$& 28 & 23&17.8& yes&\\
  09& 300.53& $+$76.20&  14.5 &  11.136 & 1.470&        &     & 0 &  $-$22 & 0.08 &$-7.15$& 32 & 45&19.3& yes&\\
  10& 333.93& $+$57.32&  15.5 &  11.327 & 1.588& var 2r &     &24 &  +43   & 0.14 &$-7.35$& 49 & 54&19.7& yes&\\
  11& 319.88& $+$30.23&  19.8 &  11.798 & 2.779& var    &     & 0 & +145   & 0.13 &$-7.30$&  - & 66&20.1&  no&\\
  12& 351.63& $+$44.74&  16.8 &  11.506 & 2.065&        &     & 0 &  +85   & 0.22 &$-7.60$& 86 & 65&20.1& yes&\\
  13& 348.10& $+$36.43&  17.1 &  10.785 & 1.809&        & H$\alpha$  & 0 & +108   & 0.23 &$-7.60$& 99 & 47&19.4& yes&\\
  14&  32.33& $+$46.37&  14.6 &  10.966 & 1.303&        &     & 0 &  +68   & 0.11 &$-6.75$& 33 & 35&18.7&  no&\\
  15& 100.83& $+$32.41&  13.9 &   9.048 & 2.503&        &     & 0 & +158   & 0.10 &$-7.35$& 24 & 19&17.4&  no&\\ 
  16&   4.40& $-$25.06&  16.7 &  10.068 & 2.565& var    & H$\alpha$  & 0 & +135   & 0.71 &$-7.35$& 66 & 29&18.4& yes&\\
  17&   7.70& $-$24.13&  14.7 &   9.981 & 1.986& var 2r & H$\alpha$  &13 & +129   & 0.38 &$-7.65$& 30 & 33&18.6& yes&\\
  18&   9.43& $-$25.08&  17.7 &   9.862 & 3.136&        &     & 7 & +132   & 0.46 &$-7.15$&117 & 25&18.0& yes&\\
  19&   1.52& $-$28.07&  13.9 &   9.244 & 2.048& var 2r & H$\alpha$  & 0 & +165   & 0.21 &$-7.60$& 23 & 23&17.8& yes&\\
  20&  19.07& $-$27.92&  11.7 &   8.109 & 1.432&        &     &15 & $-$177 & 0.36 &$-7.00$&  8 & 10&16.0& no :&\\
  21&  29.05& $-$26.26&  14.7 &   8.711 & 3.138&        &     & 0 & +55    & 0.21 &$-7.10$& 33 & 14&16.7&  no&\\
  22&  16.76& $-$38.23&  14.3 &  10.858 & 1.549& var    &     &16 & +56    & 0.28 &$-7.25$& 27 & 41&19.1& yes&\\
  23&  60.31& $-$41.67&  14.3 &   8.714 & 1.995& var    &     & 0 & $-$34  & 0.19 &$-7.60$& 28 & 18&17.3&  no&\\
  24&  26.55& $-$53.17&  15.1 &   8.922 & 2.012&        & H$\alpha$  &10 & +9     & 0.10 &$-7.55$& 42 & 20&17.5& yes&\\
  25&  25.64& $-$55.64&  15.4 &   8.882 & 2.174&        & H$\alpha$  &12 & +9     & 0.06 &$-7.50$& 48 & 19&17.4& yes&\\
  26&  35.54& $-$68.63&  15.5 &  12.280 & 1.470& var 2r & H$\alpha$  & 8 & $-$4   & 0.07 &$-7.15$& 51 & 78&20.5& yes :&\\
  27&  49.28& $-$67.38&  14.8 &   9.958 & 1.519& var 2r &     & 0 & $-$27  & 0.08 &$-7.25$& 37 & 27&18.2&  no :&\\
  28&  18.64& $-$70.94&  14.6 &  11.029 & 2.380&        &     & 0 & +94    & 0.05 &$-7.40$& 34 & 48&19.4& yes&\\
  \noalign{\smallskip}
  29& 287.82& $-$83.24&  20.1 &  11.591 & 3.286& var    &     & 0 &  -     & 0.05 &$-6.90$&  - & 50&-& Scu&\\
  30& 237.84& $-$65.37&  18.3 &  12.682 & 1.763& var    &     & 0 & +40    & 0.06 &$-7.55$& 185&112&-& For&\\

   \noalign{\smallskip}
   \hline
\end{tabular}
\end{flushleft}
\end{table*}

\section{Discussion}

\subsection{ Are there dwarf carbon stars in our sample?}

One of the major issues is to estimate distances for our C stars, and
 the main problem is to know whether they are the distant evolved AGB 
stars which were targeted when defining our selection criteria, or if some
of them are  cool dwarf carbon stars with much lower luminosity, 
as is suggested at first sight by
the surprising fact that 8 have measurable proper motions.

Therefore, it is first useful to compare the properties of our sample to the
population of dwarf carbon (dC) stars. Lowrance et al. (2003) gives an exhaustive
list of the 31 dCs presently known and considers their 2MASS $JHK_s$ data (from the
all sky release). They found that 20 dCs out of a total of 31 are detected
by 2MASS, the undetected ones being too faint. Considering the $K_s$ magnitude,
one finds that  the majority (15 dCs over 20 detected) have $K_s > 12$, and
the brightest one is at $K_s=10.48$. As for the $J-K_s$ colour, all have 
$J-K_s < 1.5$, and only 4 of 20 have $1.3 < J-K_s < 1.45$. In contrast,
all our new C stars have $K_s < 12$ with the
exception of Object \#26 (we ignore \#30 in Fornax) and more than half
of the objects (17 over 28) are brighter than $K_s=10.5$; concerning the colour,
only 4 of our total of 28 are bluer than $J-K_s = 1.45$. Therefore, when one
considers the photometric properties, the sample of our C stars is
globally very different from the sample of known dCs. 

Concerning the proper motions, we can also compare  known dCs and our C 
star sample. We retrieved the USNO-B1.0 data for the 31 dCs, and plotted them in 
Fig.~2. For 2 dCs (PG\,0824+289B and WIE93 2048-348), USNO-B1.0
gives null proper motions. For the first object, we adopted
 $\mu_{\alpha} = -28.2 \pm 1.4$\,mas\,yr$^{-1}$, $\mu_{\delta}= 0$ from Heber et al. 
 (1993), and for the second we adopted $\mu_{\alpha} = 15 \pm 24$\,mas\,yr$^{-1}$, 
$\mu_{\delta}= -3 \pm 24$\,mas\,yr$^{-1}$ from Warren et al. (1993). More
importantly, {\it there are 7 dCs which lie outside of the diagram},
because their $\mu_{\alpha}$ or $\mu_{\delta}$ are larger than 100\,mas\,yr$^{-1}$ in 
absolute value. We also plotted the USNO-B1.0 data 
for our C stars, with 20 lying at the (0,0) coordinates (null proper motion) 
and 8 at not null proper motion. It can be seen that all dCs known have
a larger $\mu$ than all our C stars except \#10 . If these C stars were in majority
dCs, one would expect larger proper motions, since they are 
in majority brighter than the known ones and would be statistically 
closer to us.

In a complementary way, instead of a statistical approach, one 
can examine in detail the 
8 objects with measurable USNO-B1.0 proper motion.  Three 
(\#10, \#22 and \#26) have $J-K_s \sim 1.5$ and  $K_s$ in the range 
10.8 to 12.3. These parameters are not atypical of dCs (see above) although marginally so. 
A very rough estimate of their distances is possible.
Following Lowrance et al. (2003; their Table 1, footnote), only
three carbon dwarfs have determined parallaxes from which one derives
$M_{K_s} \approx 6.3 \pm 0.3$. Although those dCs are warm and have
$J-K_s \approx 0.95$ within 0.05 mag., let us  adopt this luminosity for 
the moderately cooler Objects \#10, \#22 and \#26; thus, we  
obtain distances of 100, 80 and 160\,pc. Then, their proper motions
(24, 16 and 8\,mas\,yr$^{-1}$) imply transverse velocities of 11, 6 and 
6\,km\,s$^{-1}$ which, when compared to radial velocities of +43,+56 and 
$-$4\,km\,s$^{-1}$, seem plausible. These data are marginally
 compatible with expected usual disk dCs kinematics, but one has to note
 that none of these 3 stars show the strong Na\,{\sc i} lines
 or the C$_2$ $\lambda$\,6192 bandhead, as seen in some 
 cool dCs (Green et al. 1992).

If one similarly considers Objects \#17, \#18 and \#20, they are
either significantly brighter and/or redder than most known dCs
($J-K_s = 1.99,~ 3.14$ \& $1.43$ and  $K_s = 9.98,~ 9.86$ \& $8.11$ for these 
3 objects respectively). Adopting again $M_{K_s} \approx 6.3$ leads
to distance of 55, 52 and 23\,pc, and to transverse velocities of
3.4, 1.7 and 1.6\,km\,s$^{-1}$. The latter are found to be
much too small compared to the radial velocities of +129, +132 and $-$177\,km\,s$^{-1}$, 
 a situation which is very improbable. This casts some doubt on
the reliability of their proper motions, which are not large 
(13, 7 \& 15\,mas\,yr$^{-1}$) and have a probability of only 0.7--0.8.
One could argue that the  $M_Ks$ value of 6.3 adopted above might 
be in error, but if cooler C dwarfs (redder in $J-K_s$) have lower 
luminosity as for M-type dwarfs, distances and transverse velocities 
would be found even smaller, worsening the case for dCs. One could also
think to simply adjust the distance of these objects in such a way that
transverse and radial velocities be roughly equal. This is obtained by boosting
the distances by a factor of $\sim$ 50, corresponding to $M_{K_s}  \sim -1.2$:
this luminosity is typically  that of clump giants (Knapp et al. 2001), but
this solution has to be rejected because the $J-K_s$  colours of
our 3 objects ($> 1.4$) are much too red to be R-type clump giants (see Table 1 of
Knapp et al. 2001, and Ivanov \& Borissova 2002)

Finally the two remaining objects \# 24 and \#25 form an astonishing
pair of close twin objects, with  all their parameters
being almost equal: $J-K_s = $2.01 \& 2.18,  $K_s = $ 8.92 \& 8.88, $R= $ 15.1
\& 15.4, $B-R$ = 2.9 \& 3.0,  $v_{\rm helio} = $ +9 \& +9\,km\,s$^{-1}$. 
In addition, their angular separation in the sky is as low as 2.5$^{\circ}$. These
stars are exactly the kind of kinematically and spatially coherent objects which are tracers of 
halo streams! With these characteristics, especially the $K_s$ magnitude 
and the colours, it is extremely improbable that they are dCs, despite
their proper motion ($\mu \approx 11$\,mas\,yr$^{-1}$ {\it with a probability 
of 0.9  and $> 3 \sigma$ significance}).

The above remarks, together with consideration of $H\alpha$ emission and/or
variability being seen in 6  of the 8 objects with non zero $\mu$, lead 
us to have some  doubts on these $\mu$ measurements. These proper motions are
quite small ($< 25$ mas yr$^{-1}$) and  supplementary, independant  measurements
of $\mu$ are obviously needed to confirm them.  We defer to  a future work a deeper
investigation of the question whether distant, very red, possibly variable
 stars might  have
inaccurate $\mu$ measurements in USNO-B1.0.

 We tentatively conclude that our sample
does not contain C dwarfs. Our stars  are also too red in $J-K_s$
 to be clump giants or stars on the 
first ascending giant branch like the Hamburg/ESO objects.
Therefore, in the following sections, we shall adopt the view that
all our stars are genuine distant AGB C stars, and examine their location with respect to
the Sgr stream. 

\subsection{Distances}

  In order to estimate distances, we have assumed here that our  C stars
 are similar to AGB C stars located in the Sgr dwarf galaxy. This working hypothesis
 can certainly be criticized, but it is suggested by the fact that
 half of the high latitude cool C stars previously known belong to the
 tidal debris of this dwarf galaxy (Ibata et al. 2001a; see also Sect.\,1).
 Therefore, we adopt the AGB C stars of the Sgr galaxy as templates, and
 analyse their properties below to determine absolute magnitudes. 
 
 A first estimate of distances can be based on $R$-band magnitudes.
  Whitelock et al. (1999)
 published a list of 26 spectroscopically confirmed C stars in Sgr with 
 membership established on the basis of radial velocities.
 For all of them, we retrieved the 2MASS data together with the USNO-A2.0 
 $B$ and $R$ magnitudes. These 26 $R$ magnitudes range from 13.0 to 16.7, 
 with an average of 14.4. An estimate of the extinction to each star
 was obtained from the Schegel et al. (1998) tables, and the
 extinctions in the $R$ band, called $A_R$, range from 0.23 to 0.48\,mag.
 Then the mean absolute magnitude in $R$ can be obtained. For this purpose, 
 we adopt for Sgr a distance modulus $(m-M)_0$= 17.0, which is intermediate between
 the value of 16.9 taken by Majewski et al. (2003) and 17.18 considered
 by Whitelock et al. (1999). If one outlier star (the faintest in $R$) is excluded,
 one finds for the C stars of Sgr $M_{R} = -3.1$  with 25 objects, $\sigma = 0.64$
 (if all 26 stars are considered $M_{R} = -3.0$ and $\sigma = 0.79$).
 It can be noted that our $M_{R} = -3.1$ is 0.4 mag less luminous than $M_{R} = -3.5$
 adopted by Totten and Irwin (1998), and this last value would have implied
 distances 20\% larger. 
 
 For each halo C star, extinction in the $R$ band was 
 again found from the Schlegel et al.
 maps, and distances derived with $M_{R} = -3.1$ called $d_R$ are listed in Table\,5. 
 These distances are admittedly very crude, due to the low accuracy of
 USNO photometry ($\sim$ 0.5\,mag. error), possible variability, uncertainty
 in $M_R$, and especially the effect of circumstellar dust for the reddest
 stars with $J-K_s$ larger than $\sim$ 2.0. For information, Table 5 indicates
 the result of this distance scale for the  Fornax C star (\#30),
  which is not too red in $J-K_s$. No
  $R$ is available in USNO-A2.0, but two $R$ magnitudes are given in USNO B1.0: 
 with $R_2 = 18.2$, one finds 180\,kpc, while for $R_1 = 16.9$, one finds 96\,kpc. 
 These estimates are within 30\% of the true distance, 135\,kpc. In conclusion,
 the distances derived with $R$ magnitudes, called $d_R$ and listed in Table\,5,
 are probably not better than $\sim$ 30\% in relative accuracy.

 A second and probably surer estimate of distances is obtainable with 2MASS photometry.
 Considering again the 26 Sgr C stars of Whitelock et al. (1999) which
 all have 2MASS data, we compared their $K_s$ magnitudes to the averaged
  $K_s$ magnitudes of LMC C stars, this averaging being done over 
  several different bins in $J-K_s$. The Sgr 
 and LMC have similar mean reddening, $E_{\rm (B-V)} \sim 0.15$, which can be ignored
 in this comparison. We find that, for a given $J-K_s$, the C stars are fainter in 
 $apparent$ $K_s$ magnitudes in Sgr than in LMC by an average of 0.98 mag. 
 (26 objects,  $\sigma = 0.41$). By adopting  $(m-M)_0$ = 18.5 and 17.0 for LMC and 
 Sgr respectively, it is derived that, on average, the C stars of Sgr are less
 luminous by 0.50 mag in the $K_s$ band than the C stars in the LMC.
 [this shift increases to 0.67 mag is one adopts, as Majewski et al. (2003),
 distance moduli of 18.55 and 16.9 for LMC and Sgr, and is roughly consistent
 with their Figure 20 where $candidate$ Sgr C stars are compared to $candidate$ 
 LMC C stars mean locus.]
 
  We then used for template $K_s$-band luminosity the averaged $K_s$
   magnitudes of the LMC C stars  corrected
 by the above 0.50 mag. In Table\,5 are explicitly listed the adopted $M_{Ks}$ for each
 program star, and the inferred distances called $d_K$.
 These distances based on $JK_s$ are presumably more accurate than those based
 on $R$ magnitudes, because of better photometric quality,
 smaller amplitude in $K$ due to variability and reduced sensitivity to
 dust effects. However, for a given $J-K_s$, the scatter in $K_s$ for LMC C stars is
 of the order of 0.4 to 0.5 mag (1$\sigma$). Although part of this
 dispersion is due to LMC depth and inclination effects that fairly
 cancel out when an average is taken, a natural dispersion  
 $\sim$ 0.2-0.3 mag is probably present in the LMC  C stars' $K_s$ luminosities 
 (Weinberg \& Nikolaev 2001). Taking into account possible
 variability effects for our stars (maybe $\sim$ $\pm$0.2 mag. in $Ks$) 
 and the uncertainty on the
 C stars luminosity shift between LMC and Sgr, one finds that the distances of
 our program stars derived from $JK_s$ data are probably not better  than
 $\pm 25$ percent ($\pm$ 1$\sigma$). 
 
 
 Looking at Table 5, it can be seen that distances derived from $R$  and from 
 near-infrared are often in fair agreement. There are a number of cases  where $d_R$ is
 obviously too large compared to $d_K$ because the star is particularly red 
 and is presumably embedded in dust: for stars \#13 \#16 \#18 \#21 \#24 \& \#25,
 the ratio $d_R$/$d_K$ is larger than a factor of 2 and their $J-K_s$ colours
 are 1.81, 2.56, 3.14, 3.14, 2.012 \& 2.174, respectively. For object \#11,
 no value of $d_R$ is given because $R=19.8$ would lead to an exceedingly large distance (360\,kpc):
 its $d_K$ = 66\,kpc is clearly more plausible. 
 
  In the case of \#29 (in Sculptor) and \#30 (in Fornax),
 the scale adopted above  is not necessarily applicable, but remains interesting
 to consider. It leads to distances $d_K$ of 50 kpc and 112 kpc, which are
 0.57 and 0.83 times smaller than the generally adopted distances of Sculptor and
 Fornax, $\sim$ 87 and 135 kpc, respectively. While $d_K$ for \#30 is acceptable,
 the small value of $d_K$ for \#29 suggests that our rule for NIR distances
 based on Sgr templates underestimates the $K$-band luminosity of this star
 by  $\sim$ 1.2\,mag. Its membership to Sculptor remains to be definitively established
 through a radial velocity determination that our spectrum unfortunately
 could not provide. If it actually is member, its $M_{K_s}$ is $-$8.11, which is similar to
 the most luminous C stars in LMC for the same $J-K_s$ colour: in LMC with $(m-M)_0 = 18.5$,
 we find $< M_{K_s} > = -7.42 $, $\sigma = 0.53$ for $J-K_s$= 3.3,
 and the star's $M_{K_s}$ is $+ 1.3\,\sigma$ above the mean.
 
 Finally, if we exclude the 7 stars discussed above for which $d_R/d_K > 2$ 
 and the two Sculptor and Fornax objects, one finds that the log ratio 
 $x = log_{10} (d_R/d_K) $ has
 a mean of -0.025 and a dispersion of $\sigma_x$= 0.108 (N=21 objects). 
 Adopting $d_K$ as a reference, a discrepancy of $2 \sigma$ means that 
 $R$ has a typical ``error'' by $\sim 1.1$ mag., which seems in reasonable agreement with
 the fact that USNO photometry is poor and $R$ may also be variable with a comparable
 amount. In the following, the NIR-based distances $d_K$ will be adopted as 
 the surest estimates, which we recall are based on  the Sgr C stars templates.

\begin{figure}
\resizebox{\hsize}{!}{
{\rotatebox{-90}{\includegraphics{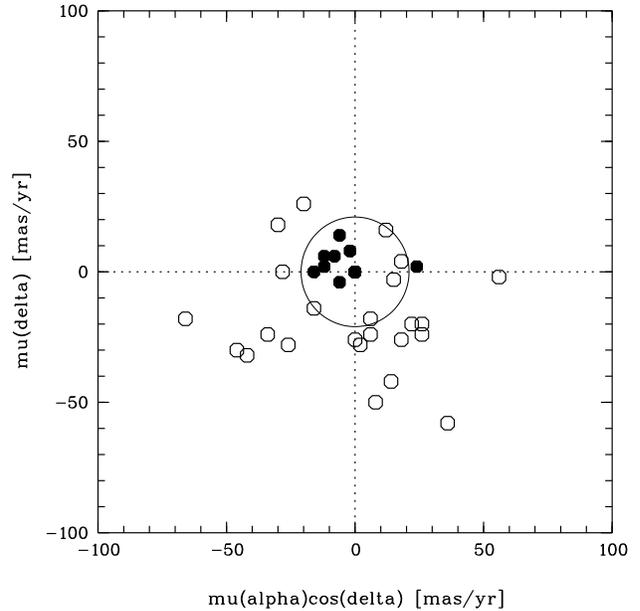}}}}
\caption[]{Proper motions of our C stars (filled octagons) as given in the
USNO-B1.0 catalogue. Twenty objects are located at the (0,0) coordinates with
no motion. Empty octagons represent the dCs presently known (USNO-B1.0 data),
but 7 dCs have too large a proper motion to be located within the limits of
this diagram. The circle represents a motion of 21 mas\,yr$^{-1}$, that was the
3$\sigma$ upper limit of TIW, enclosing 48 of the 50 APM C stars studied
by these authors.
}
\label{vfig02}
\end{figure}

\begin{figure*}
\resizebox{\hsize}{!}{
{\rotatebox{-90}{\includegraphics{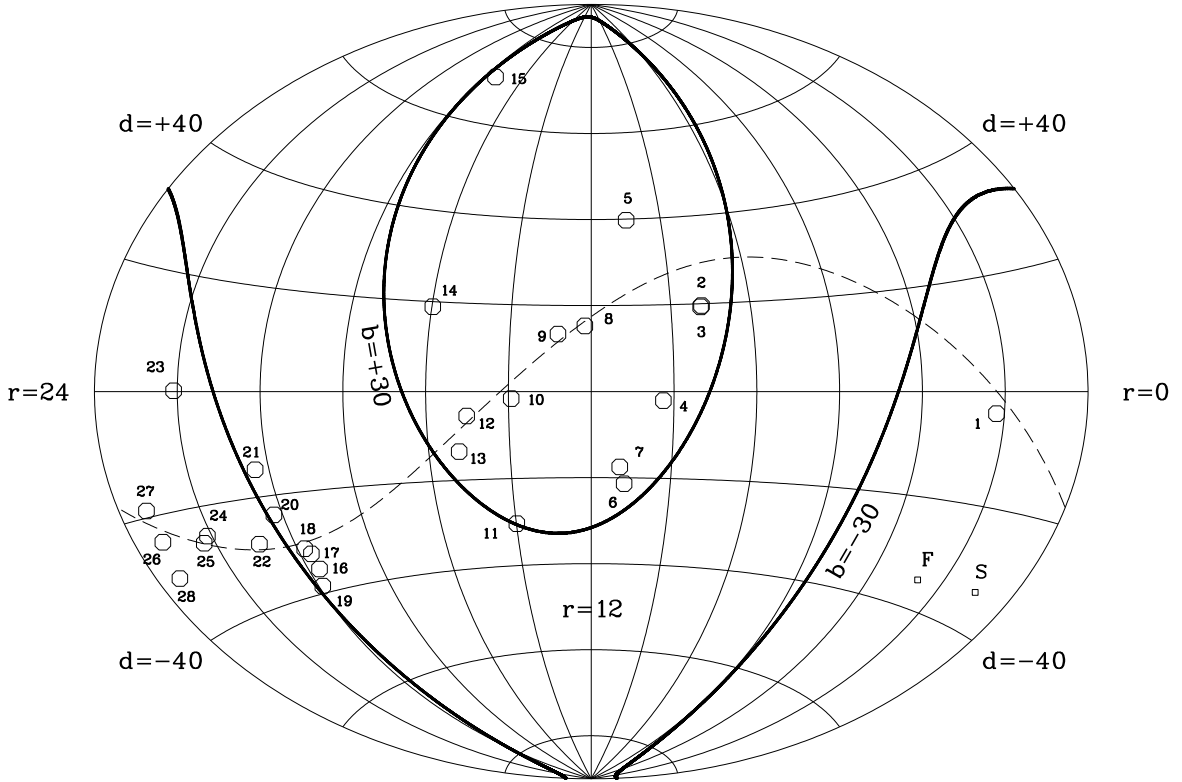}}}}
\caption[]{Aitoff map showing the location of objects in the sky. The
dashed line is a great circle with pole at galactic 
coordinates $l$ = 274$^{\circ}$, $b=-14^{\circ}$.
(Majewski et al. 2003) and this very schematically represents the Sgr orbit 
(see Fig.\,8 of Ibata et al. 2001b for a much more detailed view).
 All halo C stars have been
labelled  with their ranks as in Table 2. The last two objects (\#29 and \#30)
in the direction of Sculptor and in Fornax are the small squares with labels
S and F, respectively, seen at lower right of the map.}
\label{vfig03}
\end{figure*}

\subsection{ Location with respect to the Sagittarius Stream}

 With  positions, heliocentric radial velocities and distance estimates in hand, 
 we can examine the likelihood of association of each star with the Sgr steam. Whereas
 the path in the sky is relatively well known, especially thanks to the
 recent analysis of Majewski et al. (2003) who employ 2MASS M-type giants as tracers, 
 an accurate determination of the distances and kinematics of 
 this stream and its multiple wrapped components is not yet available. 
 Therefore, we have simply compared here our data with predictions 
 of the model by Ibata et al. (2001a) that already accounts for several observed 
 aspects of the Sgr stream and has a  (dark matter) halo density flattening 
 parameter $q_m$=0.9. 
 
  The Fig.~1 of Ibata et al. (2001a) displays in two Aitoff 
 projection maps the colour-coded  heliocentric velocities and distances of
 a Sgr stream simulation. Their distances are coded as the SLOAN apparent $g^{*}$ 
 magnitudes of A-type stars, for which they  adopted 
  $M_{g^*}$= 1.0 (these A stars are a mixture of blue horizontal branch 
  and blue stragglers; see
  Ibata et al. 2001b for more details).
 Using this absolute magnitude and the NIR distances of Table 5, we derived
 a corresponding $g^{*}_{\rm A-type}$ for each of our program stars. This $g^{*}_{\rm A-type}$ 
 is the apparent
 magnitude of an imaginary  A-type population if it were present at the distance of the C star.
 We then compared for each star $\alpha$, $\delta$, $v_{\rm helio}$ and $g^{*}_{\rm A-type}$
 with those of the model stream in Fig.~1 of Ibata et al. (2001a).
  This comparison, and therefore establishing the membership to the stream,  is difficult 
 for several objects, especially when i) uncertainties on distance or, equivalently, on 
 $g^{*}_{\rm A-type}$ are taken into account; and/or ii) the object is located
 in a region where the model stream presents a relatively low density of particles
 (in the $\alpha$, $\delta$, $v_{\rm helio}$ and $g^{*}_{\rm A-type}$ space).
 As a check, we also used the more recent work of Martinez-Delgado et al. (2003) by
 considering their diagrams showing $\delta$, $v_{\rm helio}$, and $d$ as a function of
 $\alpha$. Our final best
 estimates for membership  are indicated in Table 5 by ``yes'' or ``no'', with a ``:''
 sign for particularly uncertain cases.

 The result is that  of the 28 FHLC found in this survey, 15 are found 
 in the Sgr Stream, as indicated by ``yes'' in Table 5. It is also found
 that several distant C stars are not in the Sgr Stream,  like \#4,
 \#7 and \#11. Finding  about half of our sample in the Sgr stream is 
 comparable to the results of Ibata et al. (2001b) based on the APM and
 previously known C stars.  The same fraction is derived if only objects
 with $|b| > 30^{\circ}$ are considered. Fig.\,3 shows indeed that our stars are not
 randomly distributed in the high latitude caps, with for example
 no case at $\delta < -40$ degrees. 
 
  An interesting question is whether
 calculating distances of these high latitude C stars with the
  0.5-mag more luminous LMC C star templates 
 (instead of the Sgr C templates that we assumed here)
 would change the number of stars found in the Stream. 
 The exercise  described above provided us with a weak indication  
 that a smaller number of stars would be found to be Stream members. 
 But this question should advantageously be reconsidered with a larger sample, e.g.
 at least all presently known distant FHLC stars, and not only 
 the new objects presented here.
 Settling this question properly, i.e. by taking into account  quantitatively and
 statistically the various experimental and model errors, 
 is beyond the goal of this paper.
 
  Our spectroscopic survey is
 not yet complete and only involved candidates originating from the 2MASS 2nd Incremental
 Release which covers $\sim$ half of the sky (see e.g. Ibata et al. 2002), so that
 only supplementary observations will permit 
 a complete census and a  more detailed view of the spatial location of these
 near-infrared selected  AGB C stars
  in the halo.

\section{Summary and concluding remarks} 

In this work, we have described the first results of a survey for 
discovering new cool C stars in the
high latitude sky. The survey is  based on the exploitation of the 2MASS
 catalogue in its Second Incremental release.
 Candidates were first selected by requiring their  $J-H$, $H-K_s$ colours 
to be similar to those of already known N-type C stars in the halo, and
further examined by sample cleaning, e.g. through apparence  on the POSS plates
and presence in existing catalogues of galactic and extragalactic objects.
This selection process gave us a list of $\sim$ 200 best candidates, and for about
half of them  spectroscopy could be secured at ESO and OHP. We found 28 cool  
halo C stars: 27 objects are new; one was rediscovered erroneously (FBS\,1056+399).
 In addition, we also found  one new C star in Fornax, and one in Sculptor.

 The spectra of these C stars show in half of the cases H$\alpha$ in emission.
Also, about 2/3 of our objects could be observed between 6000 and 7000\AA\, (at ESO),
and for all of them, the spectral energy distribution is clearly rising toward
the red. These properties suggest that we are finding, 
at least in a large proportion, C stars with a N-type classification, i.e.
{\it luminous} pulsating AGB objects.

Radial velocities could be determined by cross-correlation with
templates observed with the same instrumentation, yielding velocities
accurate to $\sim$ 12 km\,s$^{-1}$ (1$\sigma$). 

A surprising fact was to find that 8 of these presumably N-type distant
stars had small but measurable proper motion measurements in the recent USNO B1.0 catalogue.
After analysing the properties of our sample, with either a statistical
approach or through consideration of individual objects, we came to the conclusion
that these  proper motion measurements are very intriguing and that the 
studied objects are  much more probably true distant AGB stars than close dCs 
with  unusual brightnesses, colours and kinematics. Yet, this point
clearly deserves further study.

Under the assumption of AGB type for the totality of our sample,
the analysis of photometric data have allowed us to estimate distances 
in the range 10 to 80\,kpc from the Sun, the distance scale being
 based on the 26 Sgr C stars of Whitelock et al. (1999). Then, consideration
 of position and radial velocities resulted in ultimately finding  about
 half of our sample in the Sagittarius stream.

In the future, it would be extremely valuable
to monitor all these stars in the NIR, as has been achieved by Feast, Whitelock
and collaborators for various galactic or Local Group AGB samples 
(e.g. Whitelock et al. 2003 and refs. therein). This would permit 
 the AGB classification
of variable objects to be ascertained, and to infer very accurate distances 
for most objects by using the period-luminosity relation.

Our survey is also far from complete: only half of our 200 best candidates  were
confirmed spectroscopically, and  the 2MASS 2nd Incremental Release which we utilised
 only covers about half of the sky at $|b|$ larger than $\sim$ 25$^{\circ}$,
 and we found 27 new  halo C stars.
It will be interesting to see  if, with the same criteria, an additional
$\sim$\,80 of these cool objects can still be discovered despite the continuing
 succession of systematic surveys.
 Enlarging the sample of distant C stars may help to further clarify
 the characteristics the Sgr stream, but also to gain a clearer view
 of the ``background'' C stars population (not in Sgr stream) and its origin. 
 It will also be a challenge to search for cases at  fainter magnitudes
 and/or lower galactic latitudes. Exploring the
invaluable 2MASS database for these luminous tracers will 
increase our knowledge on the halo properties, its stellar populations and the
  merging history of  our Galaxy.



\appendix

\section{Spectra}
Spectra obtained at ESO have a domain of $5800--8400$\,\AA\,, and those made at OHP
have a domain of $5700--6600$\,\AA\,. The ordinates are fluxes in 
erg\,s$^{-1}$cm$^{-2}$\AA$^{-1}$, after dividing these fluxes by convenient factors
given in the captions. Photometric calibration was achieved with at least
 one spectrophotometric standard star observed for each run. No correction
 was made for atmospheric extinction or slit losses. The absolute scale of these fluxes
is therefore of poor accuracy, $\sim$ a factor of 2 (note also that probably
all objects are variable), but the flux scale
from wavelength to wavelength was found to be much better ($\sim$ 20\%)
by comparing spectra of stars observed several times. The strong feature at $\sim$
7600\,\AA\, is the O$_2$ telluric absorption band. Some ESO spectra are affected
by imperfect correction of CCD fringes above $\sim$ 8000\,\AA\,.

   \begin{figure*}
     \resizebox{\hsize}{!}{\rotatebox{-90}{\includegraphics{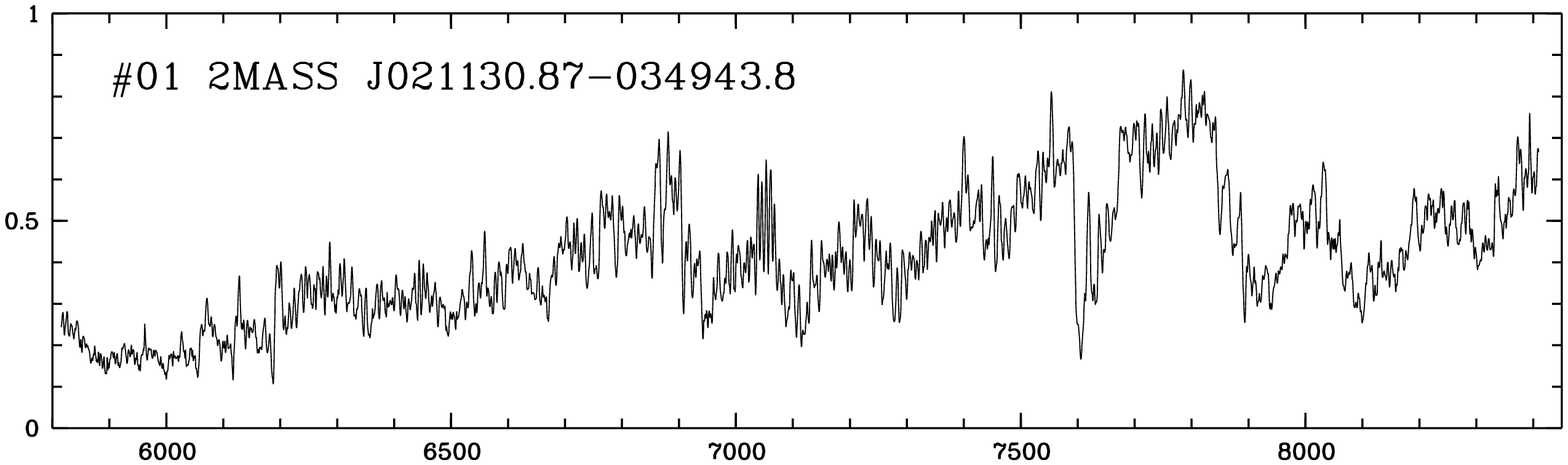}}}
     \resizebox{\hsize}{!}{\rotatebox{-90}{\includegraphics{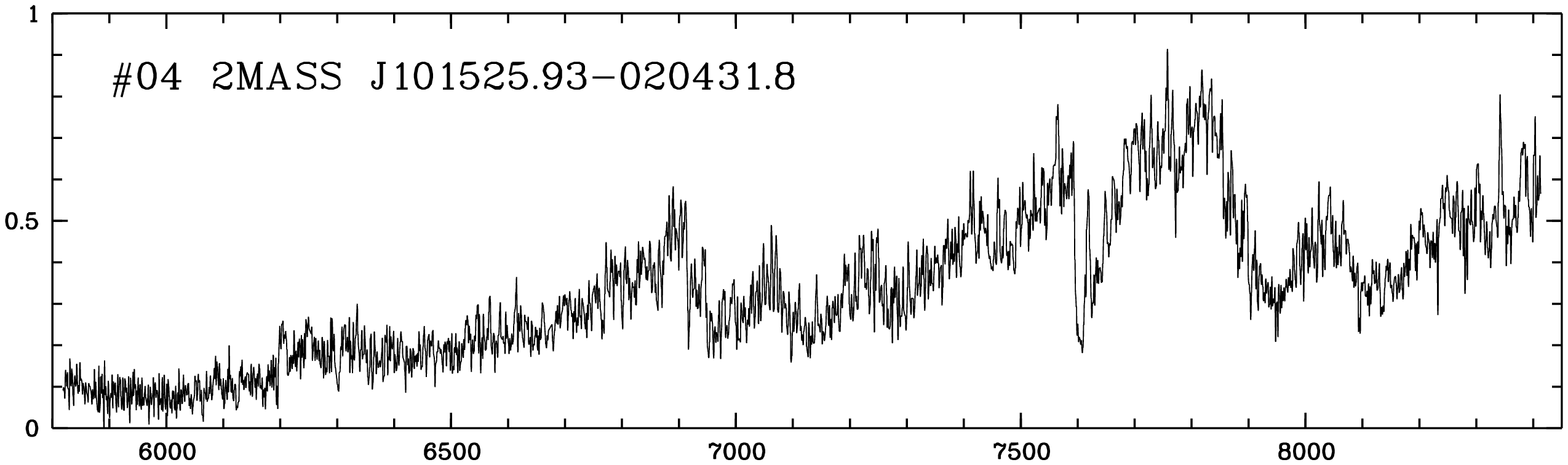}}}
     \resizebox{\hsize}{!}{\rotatebox{-90}{\includegraphics{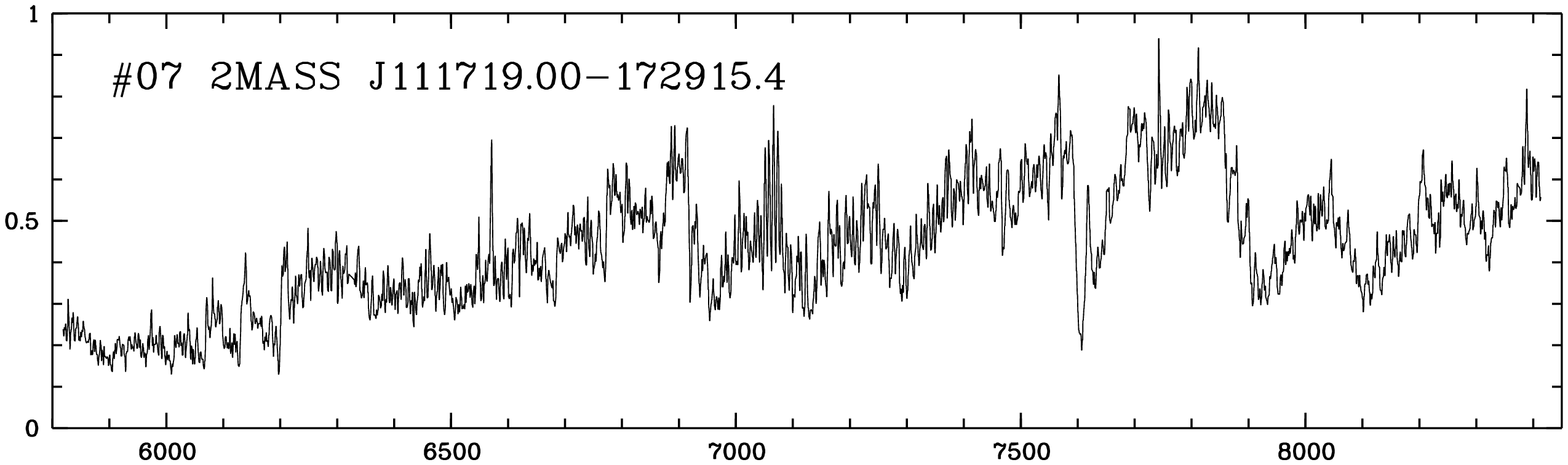}}}
    \resizebox{\hsize}{!}{\rotatebox{-90}{\includegraphics{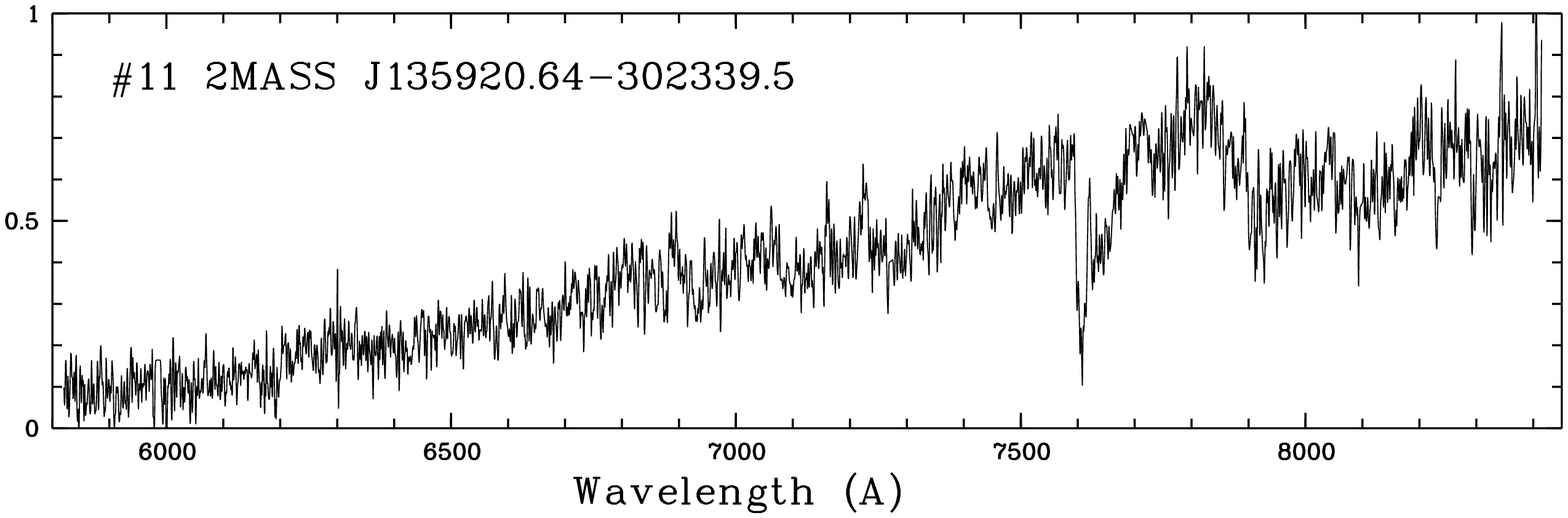}}}
     \caption[]{Spectra of objects 01, 04, 07 \& 11. In all these graphs, 
      fluxes in ordinates are in erg\,s$^{-1}$cm$^{-2}$\AA$^{-1}$. Fluxes were divided by factors
      1.5 10$^{-14}$, 1.4 10$^{-15}$,  0.9 10$^{-14}$ \& 0.6 10$^{-15}$ 
      for objects 01, 04, 07 \& 11, respectively. }
     \label{spe1}
     \end{figure*}

     \begin{figure*}
     \resizebox{\hsize}{!}{\rotatebox{-90}{\includegraphics{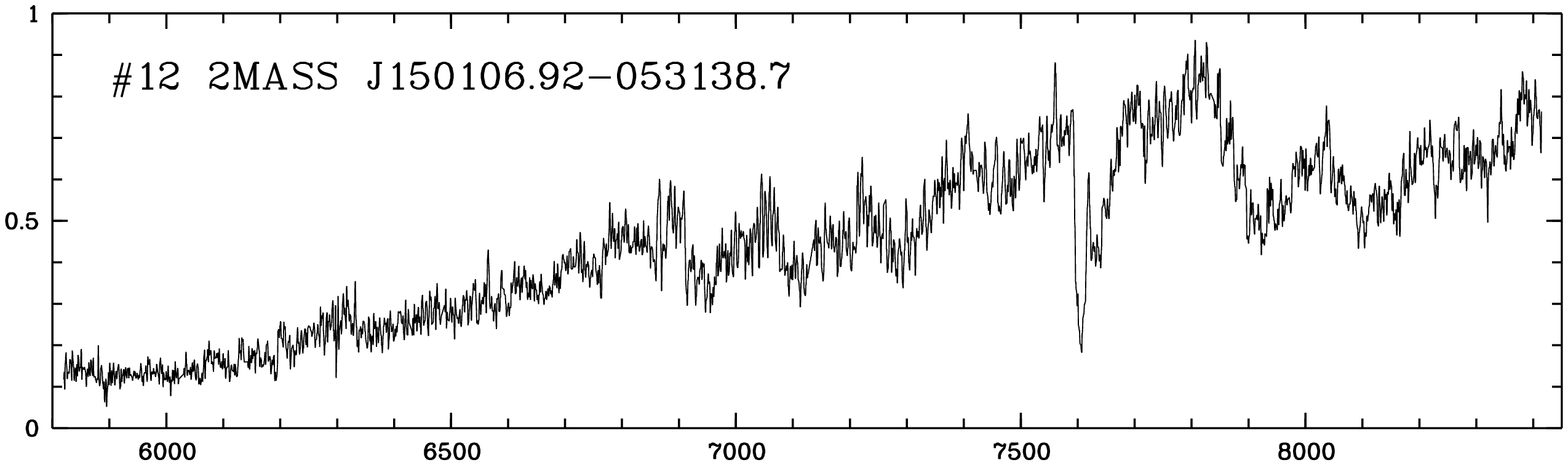}}}
     \resizebox{\hsize}{!}{\rotatebox{-90}{\includegraphics{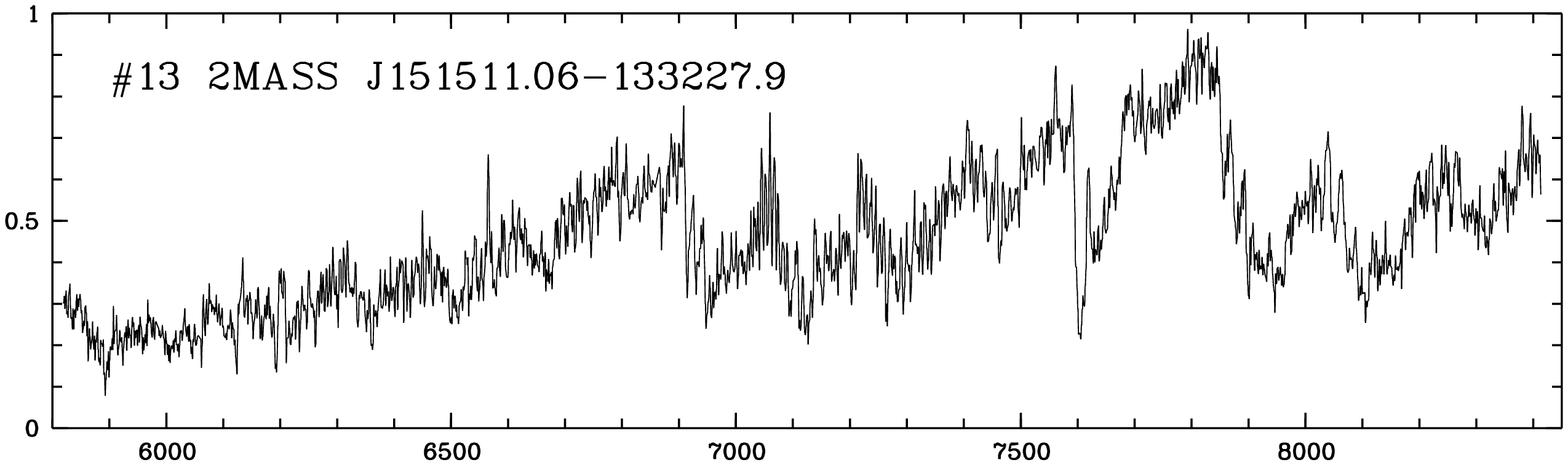}}}
     \resizebox{\hsize}{!}{\rotatebox{-90}{\includegraphics{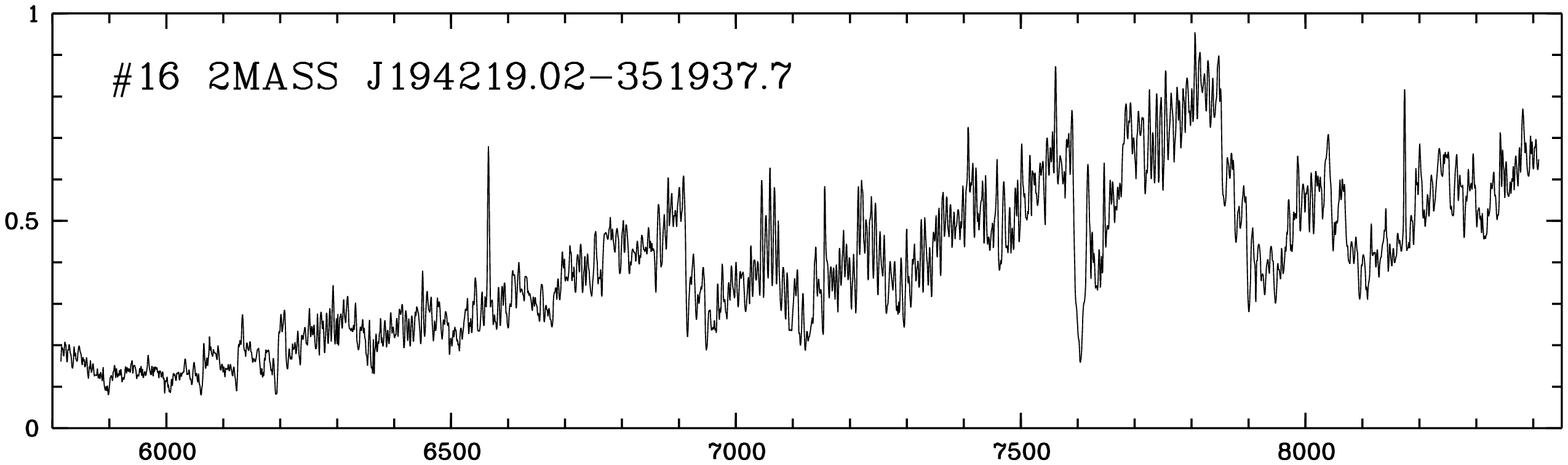}}}
     \resizebox{\hsize}{!}{\rotatebox{-90}{\includegraphics{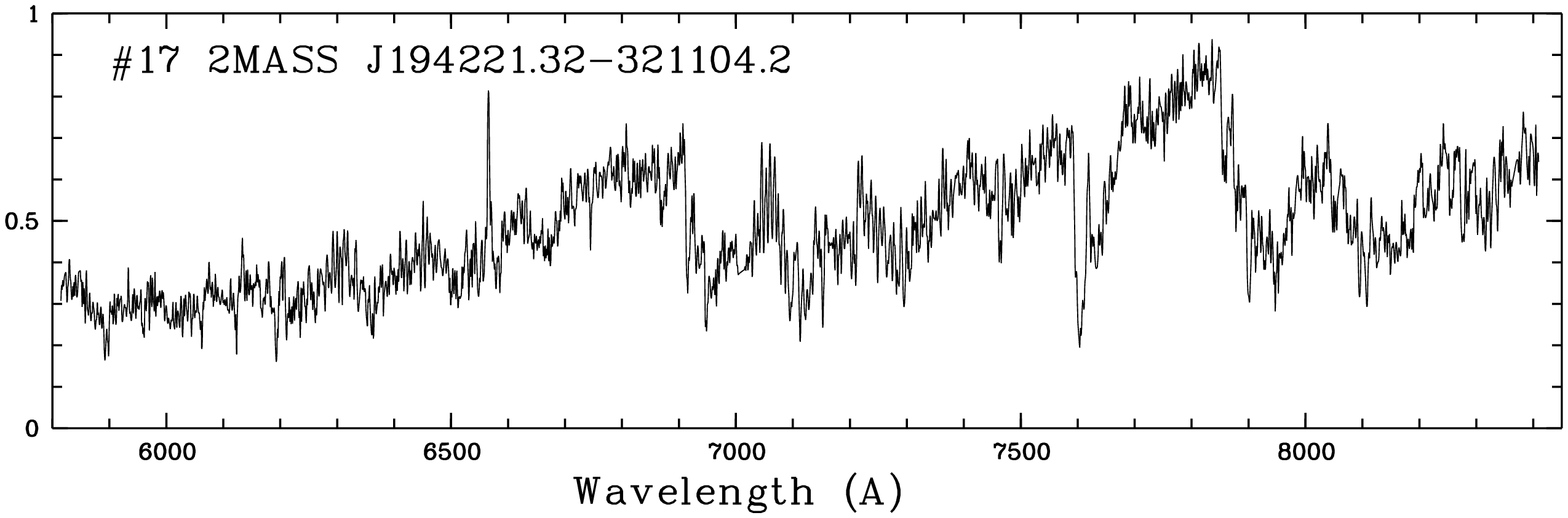}}}
     \caption[]{Spectra of objects 12, 13, 16, \& 17, for which 
      fluxes were divided by factors
      0.23 10$^{-14}$, 0.4 10$^{-14}$,  0.47 10$^{-14}$ \& 0.35 10$^{-14}$,
       respectively. }
     \label{spe2}
    \end{figure*}
    \begin{figure*}
     \resizebox{\hsize}{!}{\rotatebox{-90}{\includegraphics{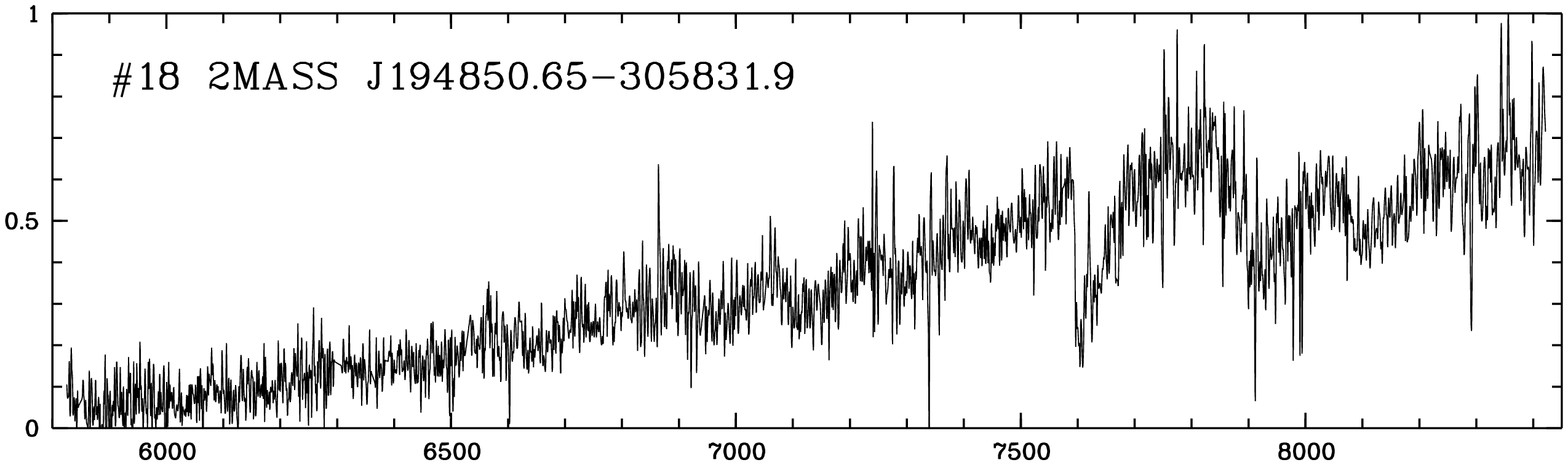}}}
     \resizebox{\hsize}{!}{\rotatebox{-90}{\includegraphics{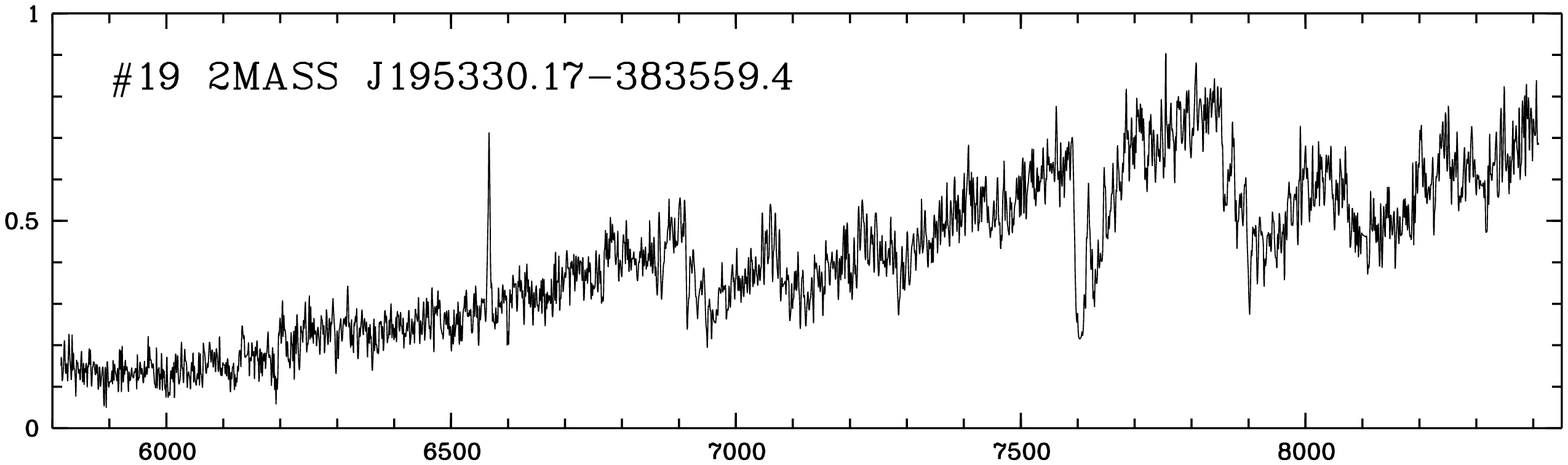}}}
     \resizebox{\hsize}{!}{\rotatebox{-90}{\includegraphics{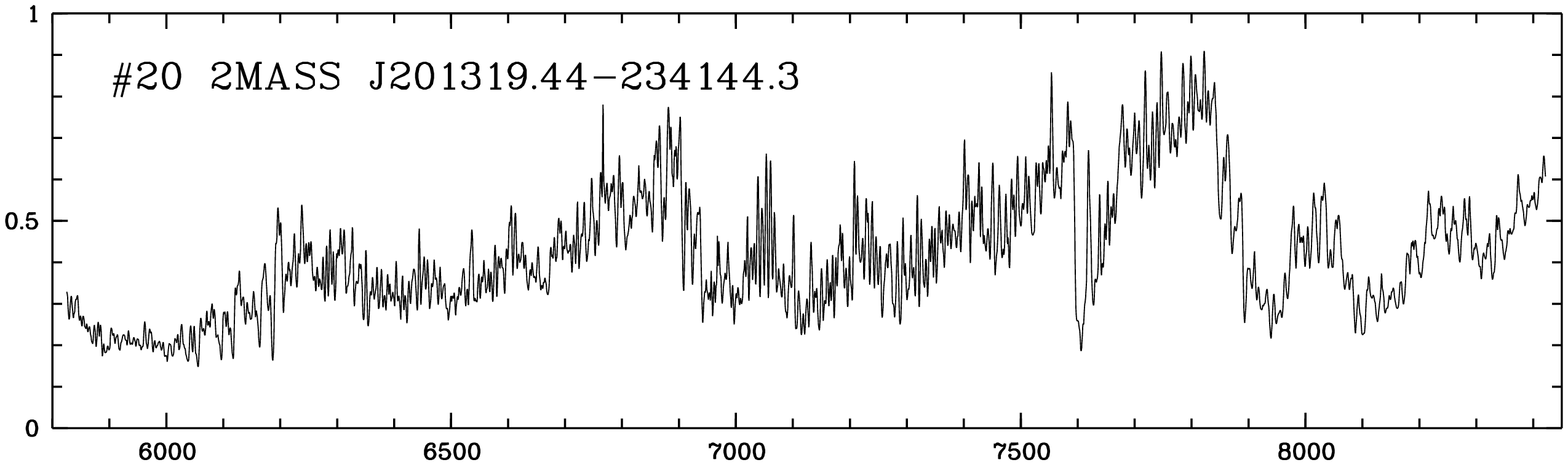}}}
    \resizebox{\hsize}{!}{\rotatebox{-90}{\includegraphics{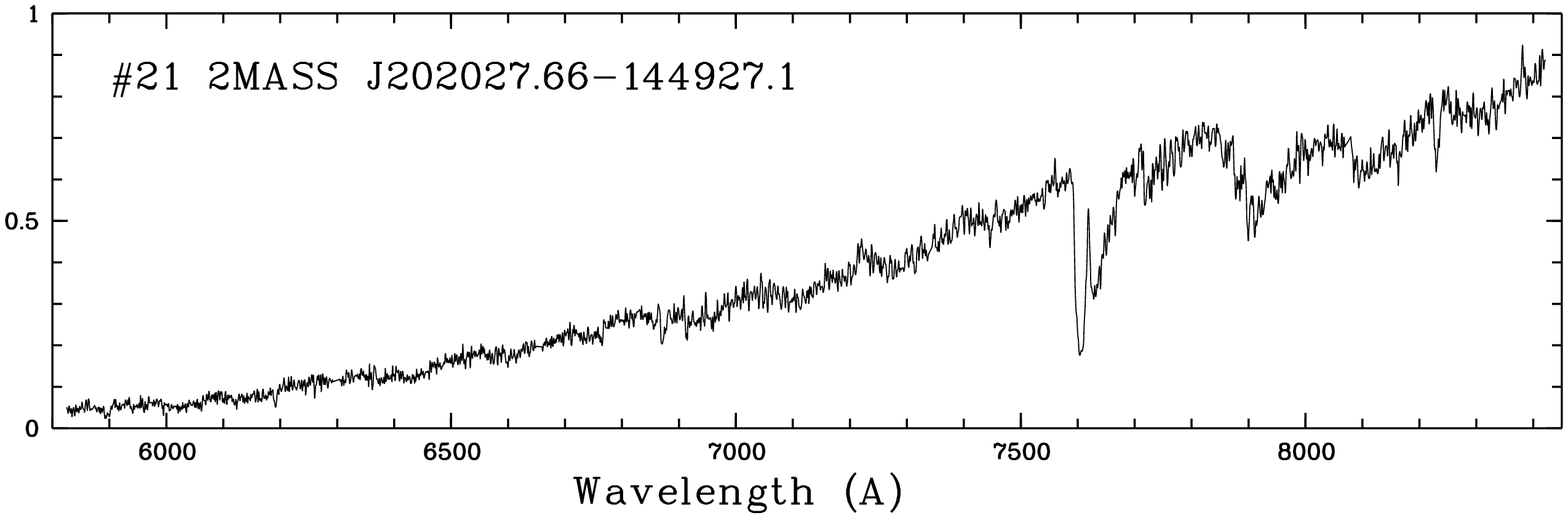}}}
     \caption[]{Spectra of objects 18, 19, 20, \& 21, for which fluxes were divided by factors
       0.22 10$^{-15}$, 0.30 10$^{-14}$,  0.60 10$^{-13}$ \& 0.27 10$^{-14}$,
        respectively. }
     \label{spe3}
     \end{figure*}
     \begin{figure*}
     \resizebox{\hsize}{!}{\rotatebox{-90}{\includegraphics{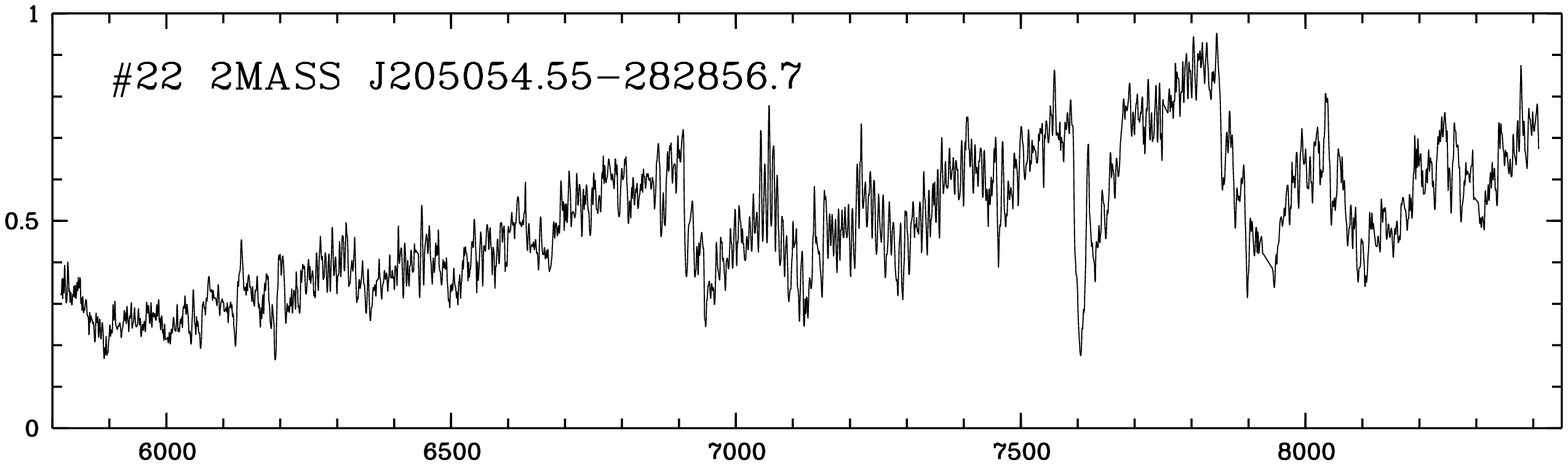}}}
     \resizebox{\hsize}{!}{\rotatebox{-90}{\includegraphics{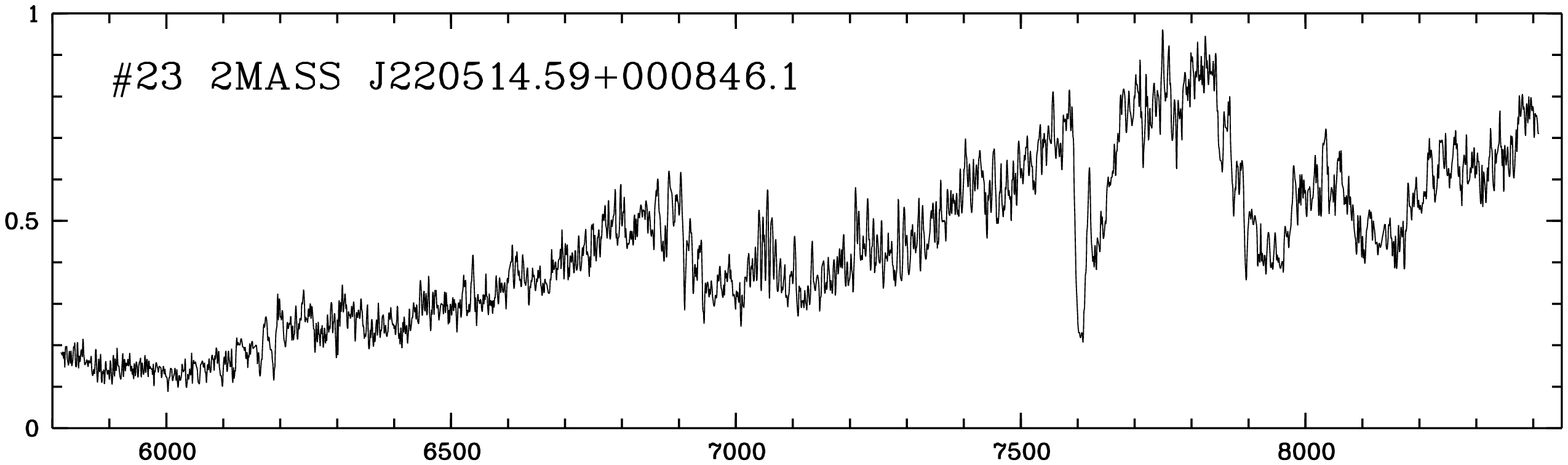}}}
    \resizebox{\hsize}{!}{\rotatebox{-90}{\includegraphics{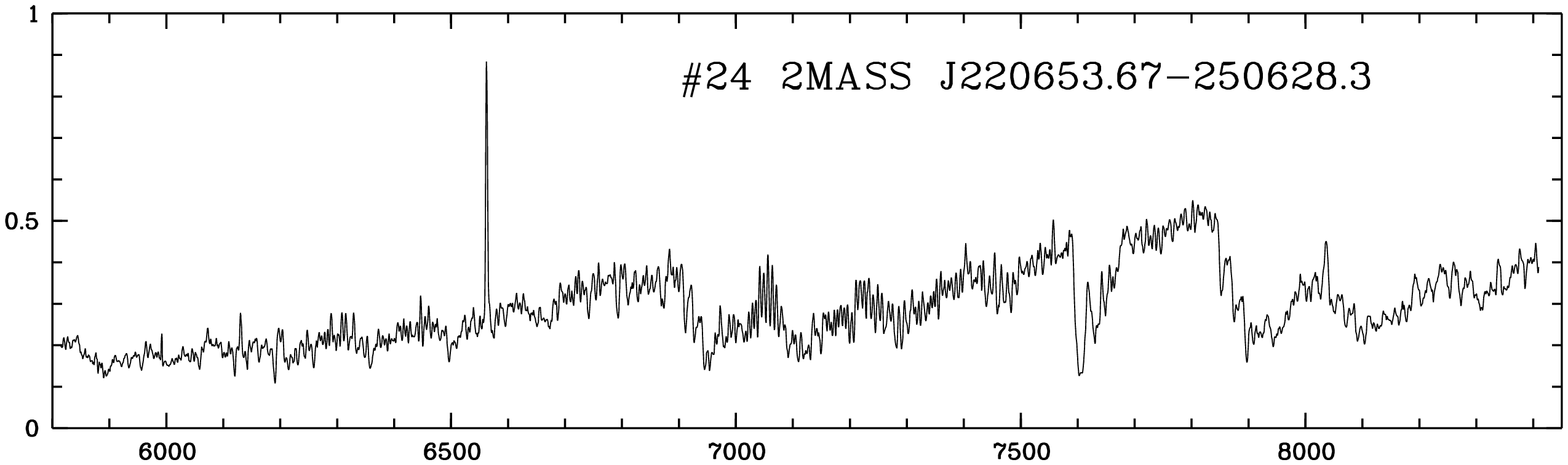}}}
    \resizebox{\hsize}{!}{\rotatebox{-90}{\includegraphics{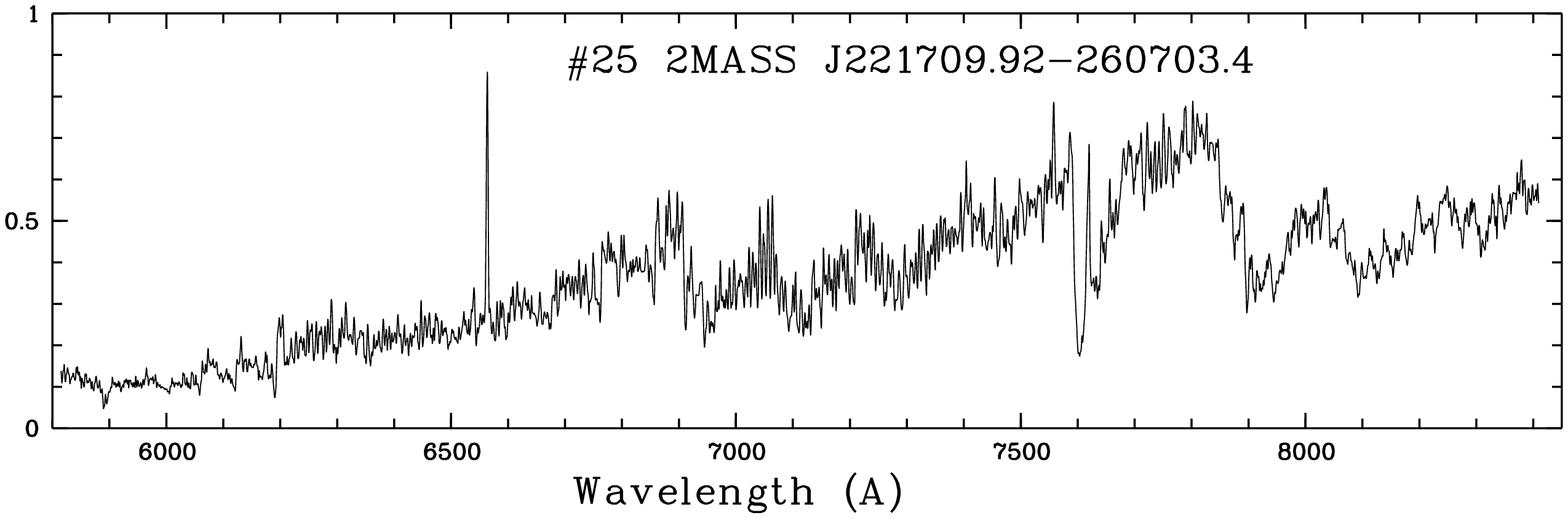}}}
     \caption[]{Spectra of objects 22, 23, 24, \& 25, for which fluxes were divided by factors
      5.5 10$^{-15}$, 6.5 10$^{-15}$,  4.5 10$^{-14}$ \& 1.1 10$^{-14}$,
       respectively. }
    \label{spe4}
     \end{figure*}
     \begin{figure*}
    \resizebox{\hsize}{!}{\rotatebox{-90}{\includegraphics{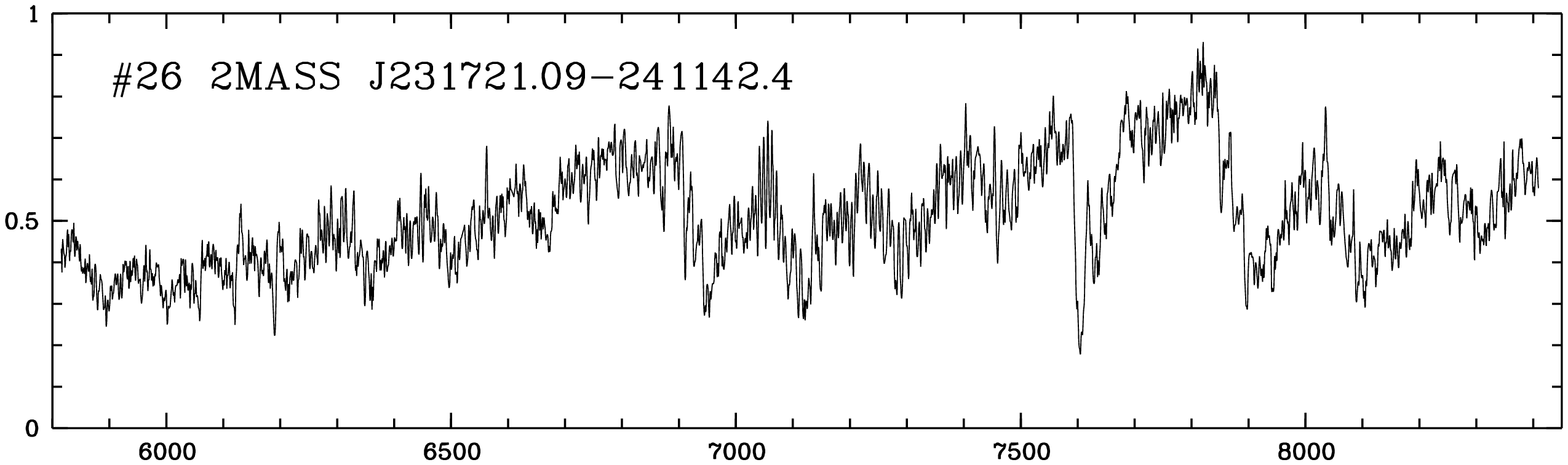}}}
    \resizebox{\hsize}{!}{\rotatebox{-90}{\includegraphics{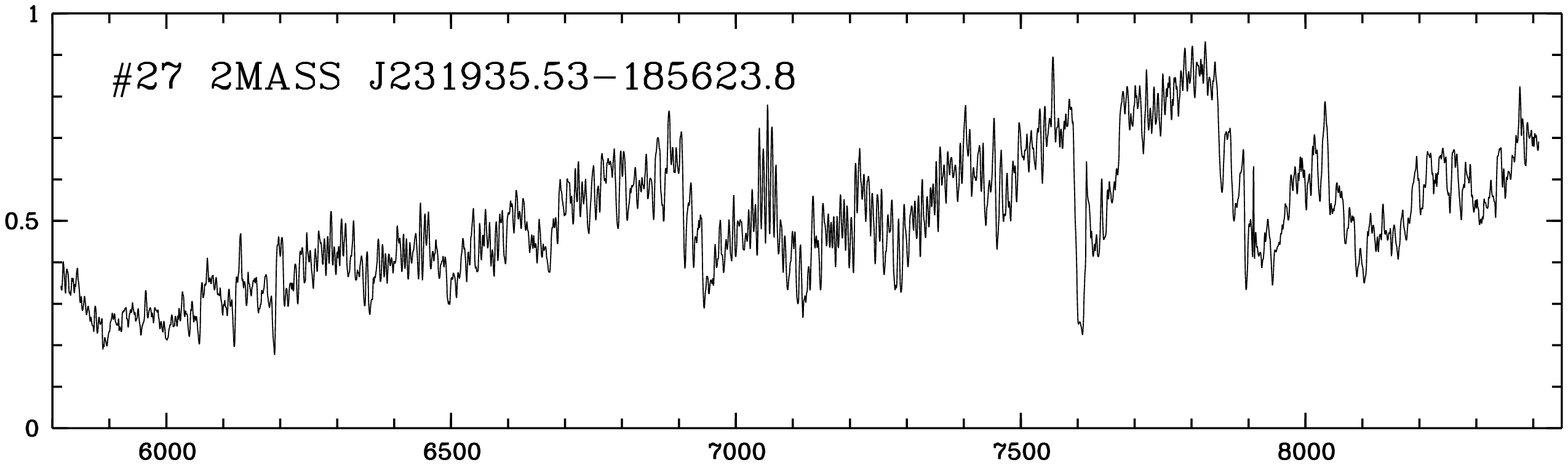}}}
     \resizebox{\hsize}{!}{\rotatebox{-90}{\includegraphics{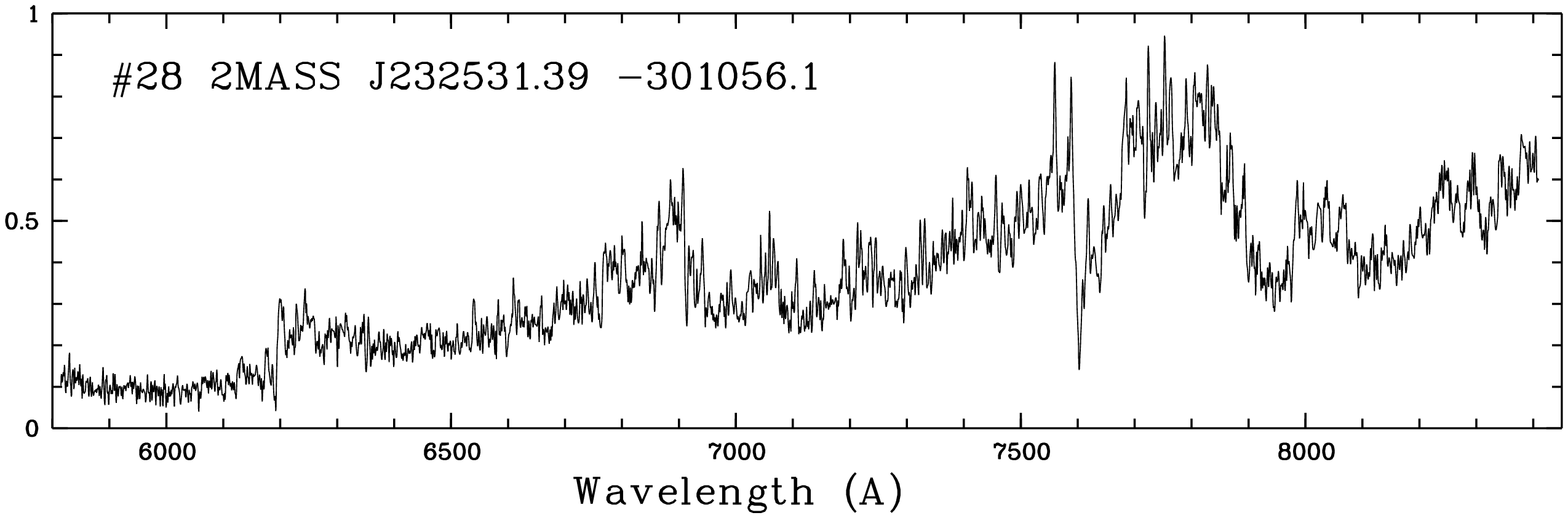}}}
     \caption[]{Spectra of objects 26, 27 \& 28, for which  
      fluxes  were divided by factors
      0.32 10$^{-14}$, 0.23 10$^{-13}$ \&  0.45 10$^{-14}$, respectively. }
     \label{spe5}
     \end{figure*}
 
      \begin{figure*}
     \resizebox{\hsize}{!}{\rotatebox{-90}{\includegraphics{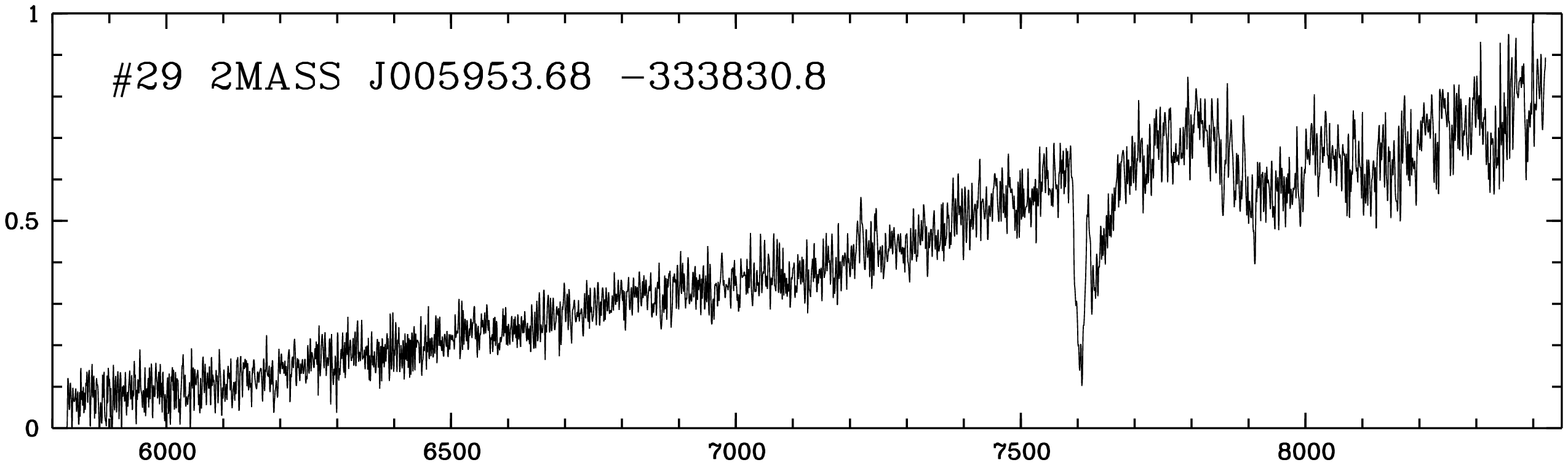}}}
     \resizebox{\hsize}{!}{\rotatebox{-90}{\includegraphics{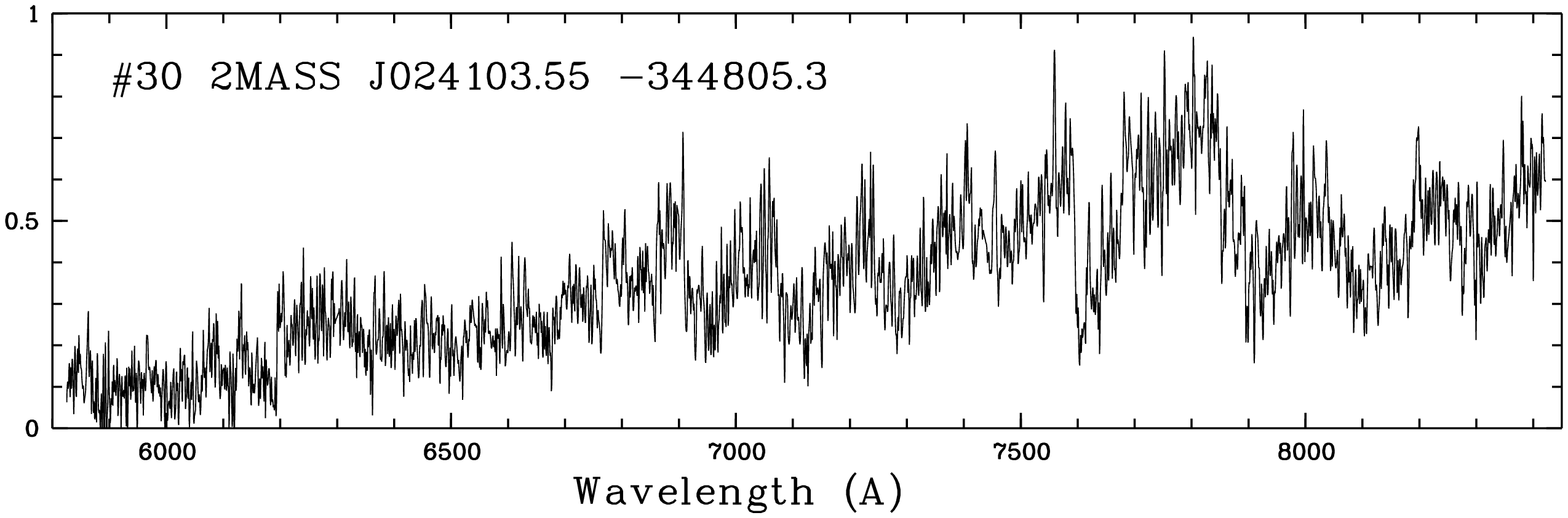}}}
     \caption[]{Spectra of objects 29 \& 30, for which 
     fluxes  were divided by factors
      0.55 10$^{-15}$, 0.40 10$^{-15}$, respectively. }
     \label{spe5b}
     \end{figure*}


     \begin{figure*}
     \resizebox{\hsize}{!}{\rotatebox{-90}{\includegraphics{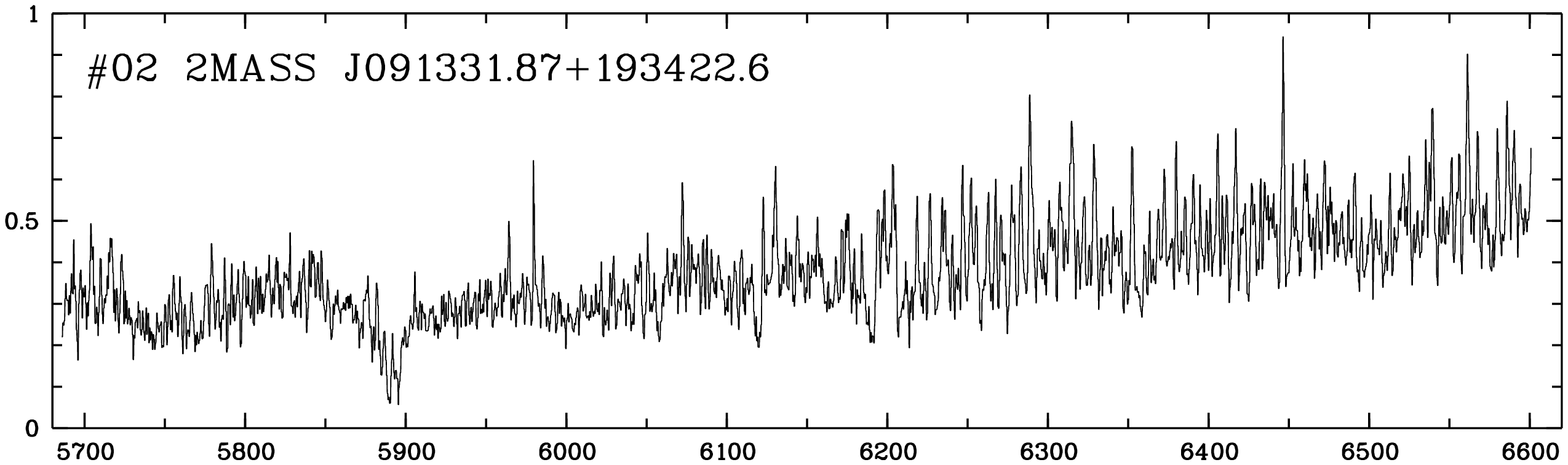}}}
     \resizebox{\hsize}{!}{\rotatebox{-90}{\includegraphics{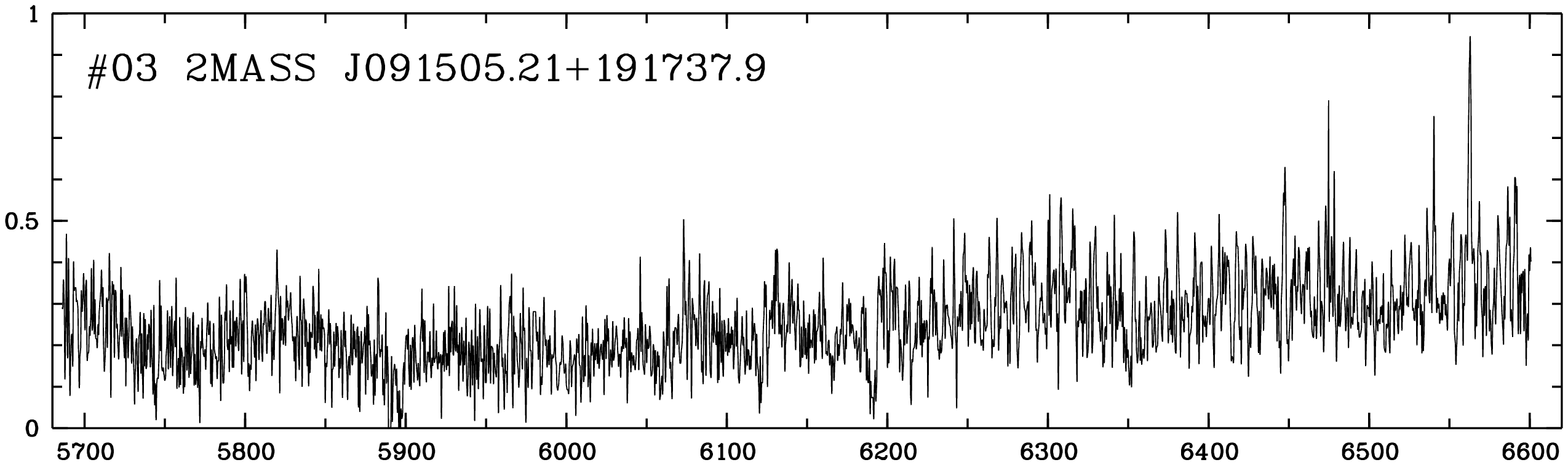}}}
     \resizebox{\hsize}{!}{\rotatebox{-90}{\includegraphics{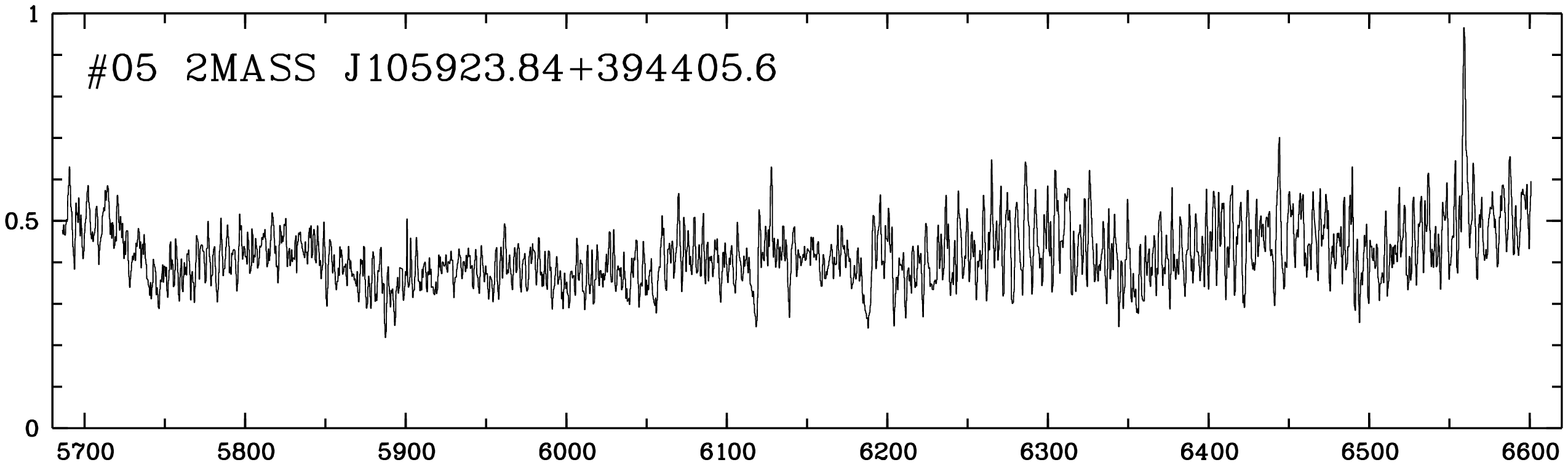}}}
     \resizebox{\hsize}{!}{\rotatebox{-90}{\includegraphics{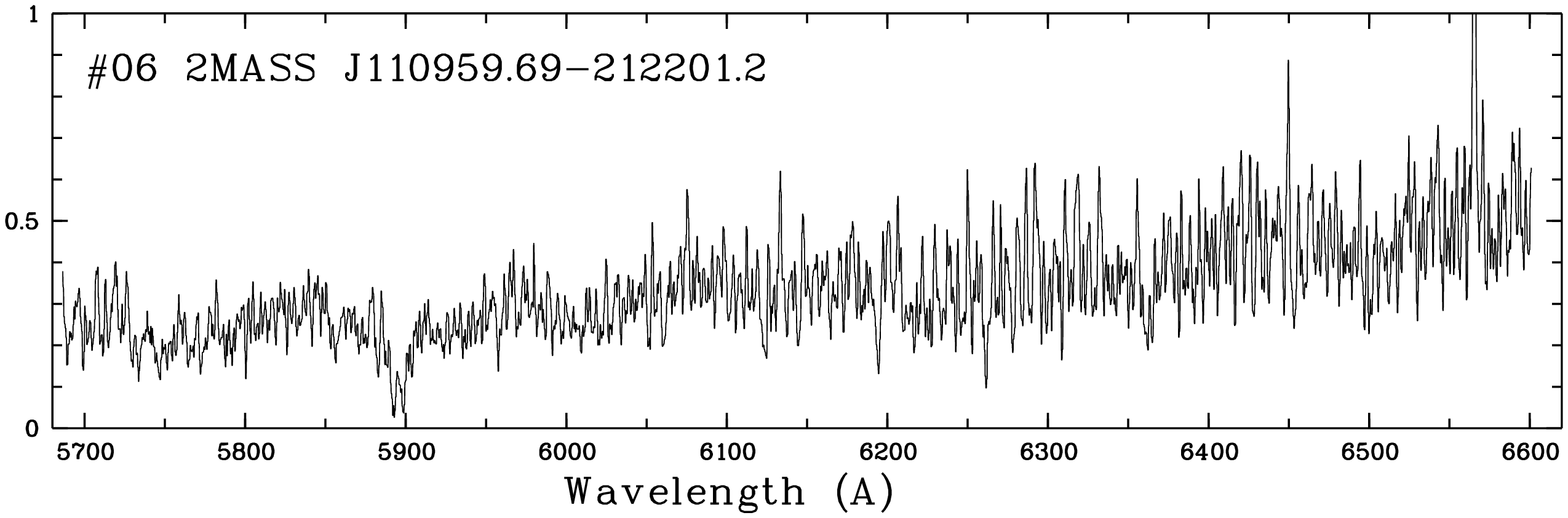}}}
     \caption[]{Spectra obtained at OHP of objects 02, 03, 05 \& 06, for which  
      fluxes in ordinates  in erg\,s$^{-1}$cm$^{-2}$\AA$^{-1}$ were divided by factors
      4.7 10$^{-14}$, 0.90 10$^{-14}$ , 0.57 10$^{-13}$  \& 0.10 10$^{-12}$,
       respectively. }
    \end{figure*}
 
     \begin{figure*}
    \resizebox{\hsize}{!}{\rotatebox{-90}{\includegraphics{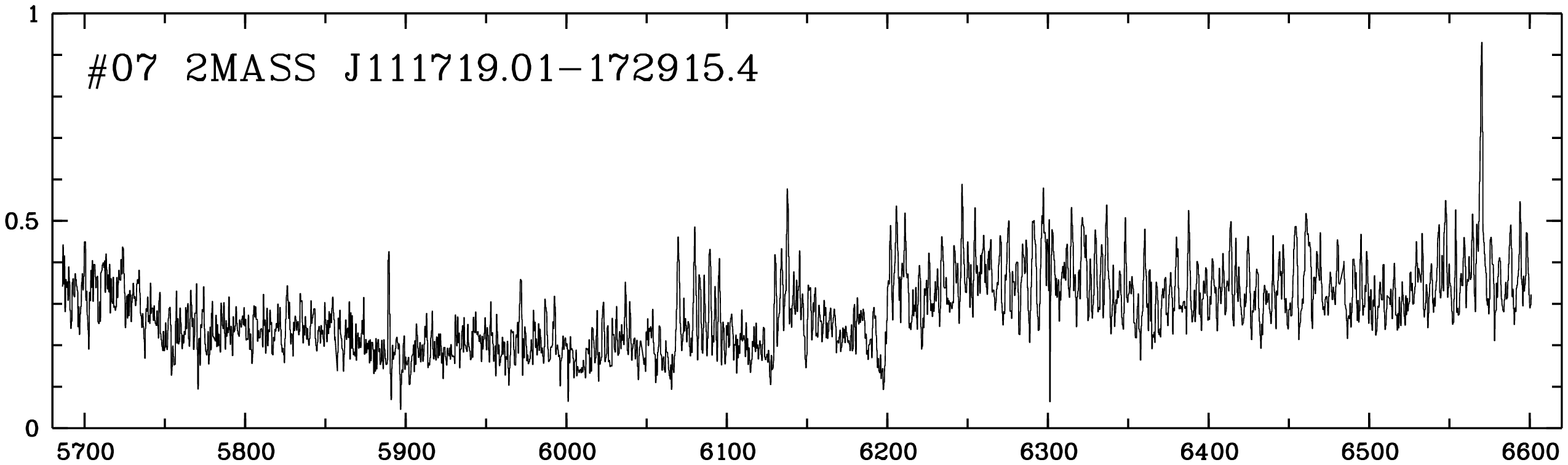}}}
     \resizebox{\hsize}{!}{\rotatebox{-90}{\includegraphics{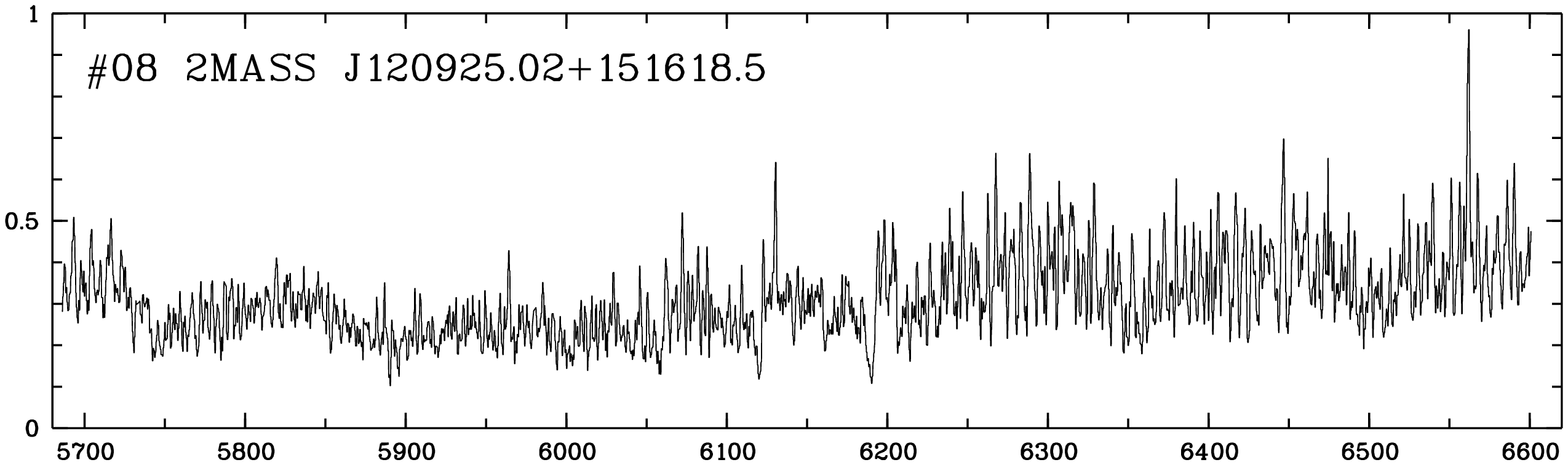}}}
     \resizebox{\hsize}{!}{\rotatebox{-90}{\includegraphics{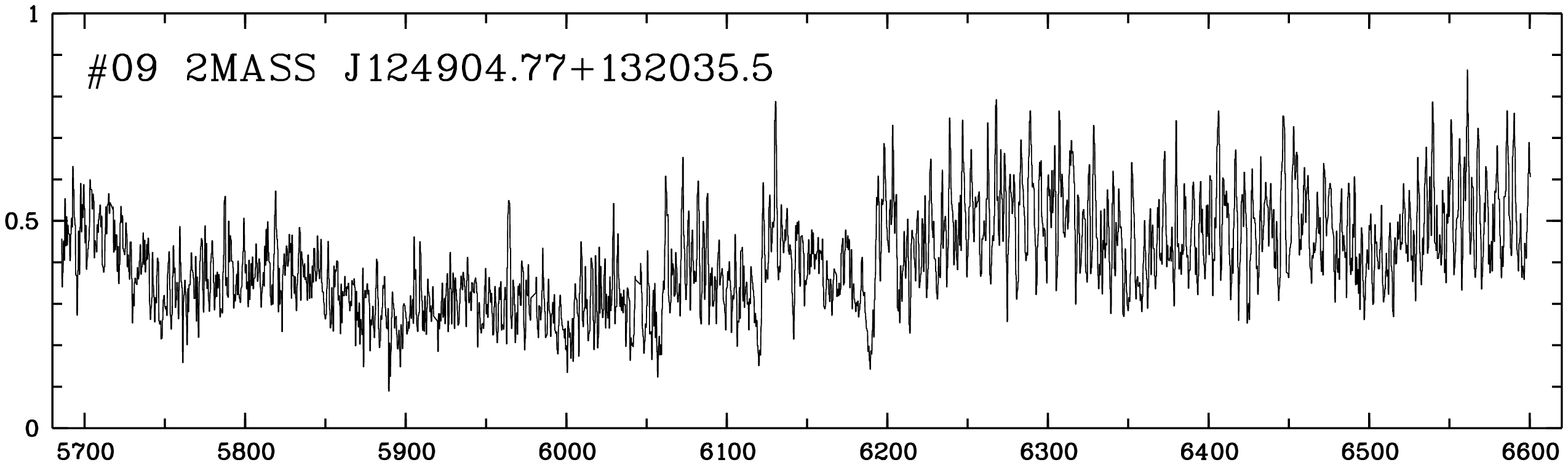}}}
     \resizebox{\hsize}{!}{\rotatebox{-90}{\includegraphics{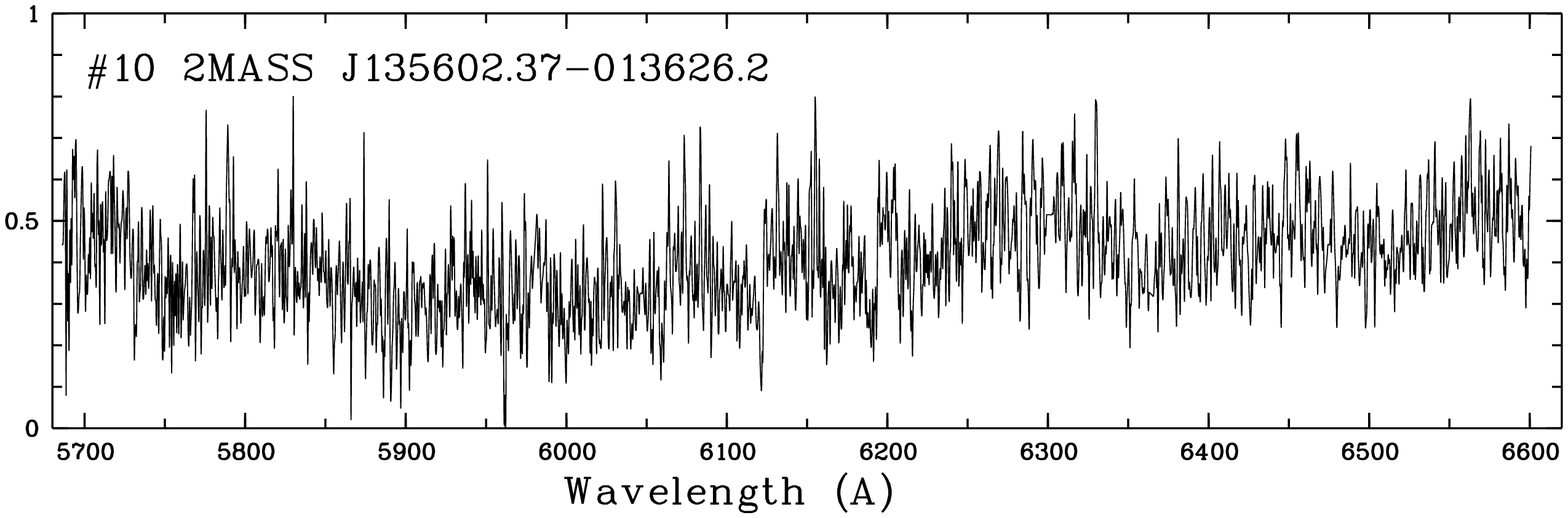}}}
     \caption[]{Spectra obtained of objects 07, 08, 09 \& 10, for which  
     fluxes  were divided by factors
     0.65 10$^{-14}$, 1.90 10$^{-14}$ , 0.70 10$^{-14}$  \& 1.2 10$^{-15}$,
      respectively. }
     \label{spe6}
     \end{figure*}

     \begin{figure*}
     \resizebox{\hsize}{!}{\rotatebox{-90}{\includegraphics{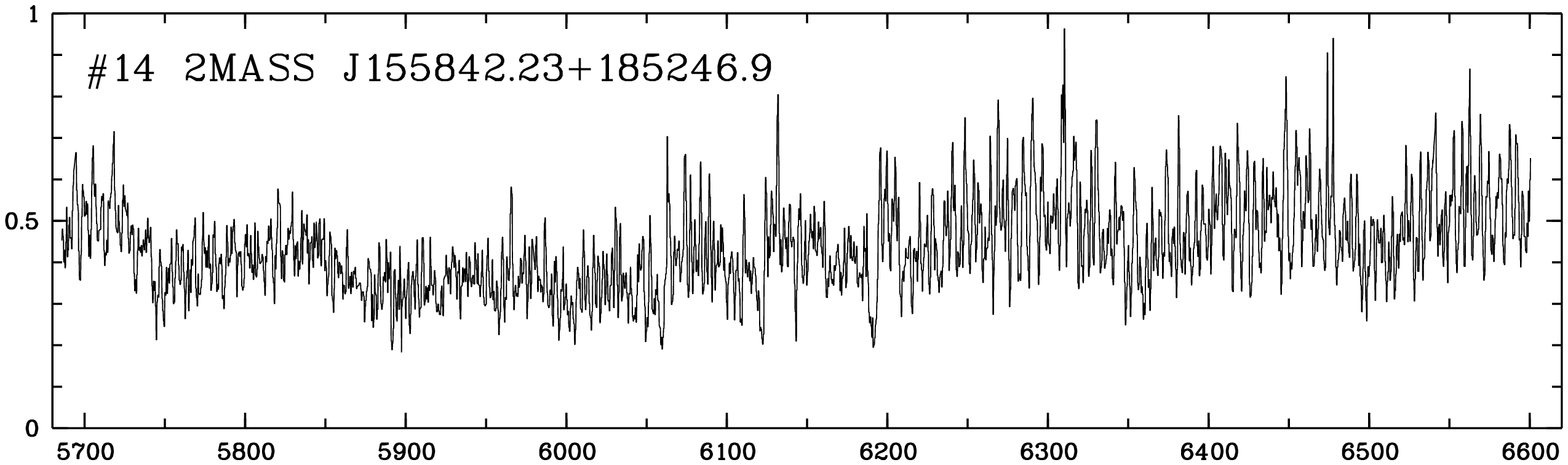}}}
     \resizebox{\hsize}{!}{\rotatebox{-90}{\includegraphics{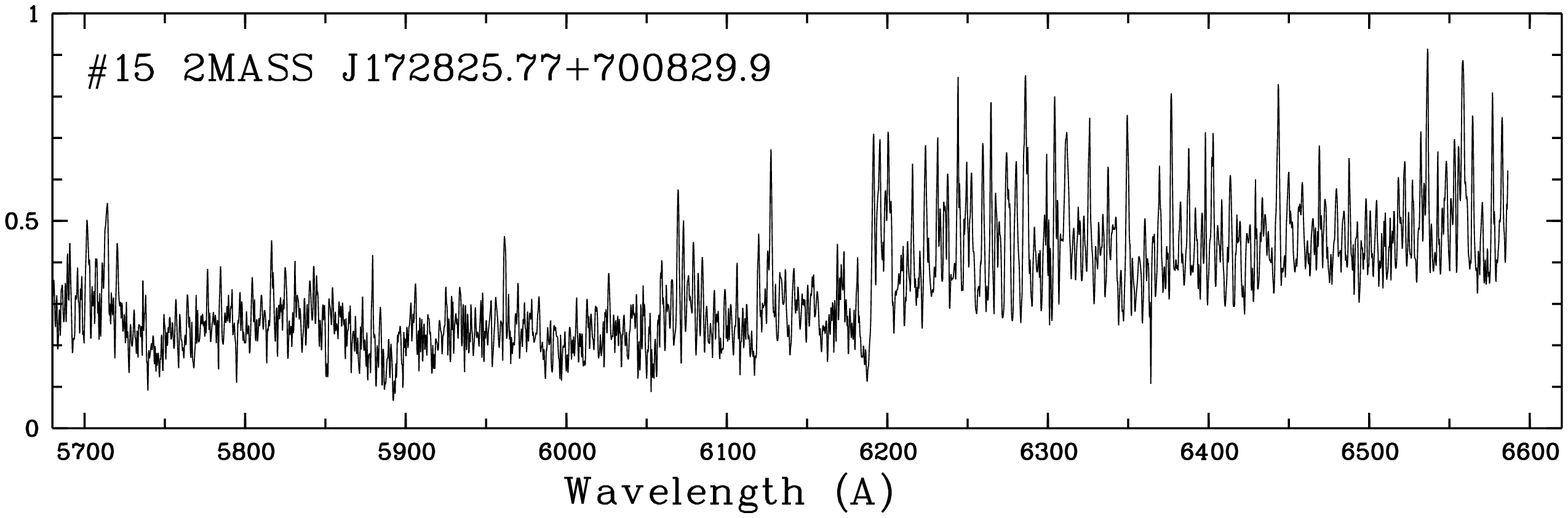}}}
     \caption[]{Spectra of objects 14  \& 15, for which 
      fluxes were divided by factors
     1.30 10$^{-14}$ \& 0.55 10$^{-14}$, respectively. }
     \label{spe7}
     \end{figure*}

\begin{acknowledgements}

The use of data products from the Two Micron All Sky Survey 
(2MASS), which is a joint project of the Univ. of Massachusetts and  the Infrared
Processing and Analysis Center / California Institute of Technology, funded
by NASA and NSF, is greatly appreciated; this program could not have been
performed, or even started, without the rapid, extensive and public distribution  
of 2MASS data.
This work also benefitted from using
the CDS database of Strasbourg, and the  impressive US Naval Observatory
 astrometric and image database. We also used the POSS-UKST
  Digitized Sky Survey made available by the Canadian Astronomical Data Center and
by the ESO/ST-ECF Center in Garching. We thank the anonymous referee for
suggestions that helped to clarify several points. We would like also to thank
the staff at OHP and ESO,  and particularly John Pritchard at La Silla.
This research was supported  by CNRS ``Programme National Galaxies'' 
(N.M.) and through
the Jumelage 18 ``Astrophysique France-Arm\'enie'' (K.G. and M.A.).

\end{acknowledgements} 


\end{document}